\documentclass[12pt,a4paper]{article}

\usepackage[utf8]{inputenc}
\usepackage{mathrsfs}
\usepackage{float}
\usepackage[T1]{fontenc}
\usepackage{amsmath,amssymb,amsfonts}
\usepackage{amsthm}
\usepackage{graphicx}
\usepackage[margin=1in]{geometry}
\usepackage{abstract}
\usepackage{multicol}
\usepackage{titling}
\usepackage{subcaption}
\usepackage{todonotes}
\usepackage{authblk}
\usepackage{geometry}
\usepackage{algorithm}
\usepackage{algpseudocode}

\graphicspath{ {./img/} }

\usepackage{bm}
\usepackage{booktabs}
\usepackage{hyperref}
\hypersetup{colorlinks=true, linkcolor=blue, urlcolor=blue}
\title{Parameter conditioned interpretable U-Net surrogate model for data-driven predictions of convection-diffusion-reaction processes} 

\author[1]{Michael Urs Lars Kastor\thanks{Corresponding authors: \href{mailto:m.kastor@rptu.de}{m.kastor@rptu.de}, \href{mailto:jan.rottmayer@rptu.de}{jan.rottmayer@rptu.de}}}
\author[1]{Jan Rottmayer$^{*}$}
\author[1]{Anna Hundertmark}
\author[1]{Nicolas Ralph Gauger}
\affil[1]{RPTU University Kaiserslautern-Landau, Germany}
\date{}
\newcommand{\R}{\mathbb{R}}
\newcommand{\norm}[1]{\left\lVert #1 \right\rVert}

\begin{document}
	
\maketitle
\begin{abstract}
    We present a combined numerical and data-driven workflow for efficient prediction of nonlinear, instationary convection-diffusion-reaction dynamics on a two-dimensional phenotypic domain, motivated by macroscopic modeling of cancer cell plasticity. A finite-difference solver, implemented in C++, is developed using second-order spatial discretizations and a step-size controlled Runge-Kutta time integrator. A mesh refinement study confirms the second-order convergence for the spatial discretizations error. Based on simulated input-output pairs and corresponding parameterizations for the diffusion, advection, and reaction mechanisms, we train a parameter-conditioned U-Net surrogate to approximate the fixed-horizon solution map. The surrogate incorporates Feature-wise Linear Modulation (FiLM) for parameter conditioning, coordinate encoding to incorporate spatial location information, and residual blocks to enable multiscale representation learning in combination with the U-Nets skip connections. The trained model achieves low prediction error on held-out test data and provides favorable prediction times due to the GPU based parallelization. Generalization is analyzed using a factorial test dataset, separating initial conditions from parameter conditioning. The results reveal that approximation difficulty varies primarily with the conditioning vector (i.e., the induced PDE regime), rather than with the initial conditions. 
\end{abstract}


\section{Introduction}
In this study, we combine a robust numerical solver and a deep neural network (DNN) surrogate model to predict the solutions of nonlinear, instationary  convection-diffusion-reaction partial differential equations (PDE) based on simulation data. The chosen PDE model can be applied to or generalized for a broad spectrum of applications, including, among others, fluid dynamics coupled to diffusion processes or chemical reactions. 
However, the investigations of this study are motivated by the macroscopic modeling of cancer cell phenotypic dynamics in terms of cell populations, for which the studied PDE model is applied. In this context, it should be emphasized that the description of the phenotypic plasticity of cancer cells (i.e., their the ability to change phenotype in response to the environment) and uncovering its underlying mechanisms remain areas of intense research activity and research interest. Many different strategies are currently applied to study this phenomenon. 
To the best of our knowledge, the key mechanisms responsible for changes in the phenotype of cancer cells, leading to therapy resistance, have not yet been fully uncovered.
This is partly due to the difficulties associated with the high dimensionality of cell transcriptomic data and the loss of biological information when working with low dimensional representations instead.

Radig et al. \cite{radig} and Colson et al. \cite{colson} each highlight a few of the many different approaches to describe cancer cell plasticity. 
\cite{radig} focuses on comparing different deep learning tools working directly on single-cell transcriptomic datasets and concludes that the description of the cell response in this lightly processed representation could be unattainable. 
This suggests that it is probably not advisable to work directly with single-cell transcriptomic data, but rather with a representation in a reduced space.
In contrast to the deep learning approaches, \cite{colson} highlights methods utilizing lower dimensional representations, e.g., classes of phenotypes or phenotype-spaces with reduced dimensionality, including approaches via stochastic and ordinary differential equations (to be found in, e.g., \cite{lee,lavenant}) and partial-(integro)-differential equations (P(I)DE), see \cite{chrisholm, browning, almeida, xiao} for more details.
In this context, we refer to the approach outlined by Chrisholm et al. \cite{chrisholm}, describing phenotypic plasticity with a PIDE model in a two-dimensional phenotypic space with survival and proliferation potential as the two dimensions. In their PIDE they account for non-genetic instability represented by a diffusion term, for selection via population dynamics represented by a nonlinear reaction term (including an integral term) and for drug-induced shift mechanisms represented by a convection term. \\
Based on a suitable low-dimensional phenotypic representation of real biological data and the macroscopic P(I)DE model under consideration, data-driven approaches for fitting the model to the processed observation data
can be applied to identify the model parameters and thus the driving mechanisms of phenotype change.

\medskip 

In this work we follow similar macroscopic modeling approach as in Chrisholm et al. \cite{chrisholm} in a two dimensional phenotypic space, but constrain the model to a convection-diffusion-reaction PDE with parametrized coefficient functions.
We develop and validate a numerical solver implemented in C++ based on the second order finite-difference method for spatial discretization and third order Runge-Kutta method with step-size-control for temporal discretization.
Although this solver provides accurate numerical results, the repeated PDE evaluations required for envised parameter identification tasks result in a high computational load. 

To mitigate this bottleneck, aiming for a fast
prediction of the PDE solutions, a data-driven surrogate model is introduced. Neural network surrogates are especially attractive because modern machine learning frameworks such as PyTorch\cite{paszke2019pytorch, pytorch2024} and JAX\cite{jax2018} provide efficient GPU/TPU based parallelism and automatic differentiation. This enables fast inference and gradient-based sensitivities for later parameter inference tasks. 
There have been several competing approaches for PDE operator surrogates using neural networks. Physics-Informed Neural Networks (PINNs)\cite{raissi2017pinns} have been proposed introducing the physics and dynamics directly into the learning objective of the neural network and have become extremely popular in recent years with various adaptations as summarized in Sophiya et al. \cite{sophiya2025pinns}. Despite their conceptual appeal, several limitations of PINNs have been documented, particularly in regimes relevant to convection–diffusion–reaction systems. First, PINNs require repeated evaluation of PDE residuals and, depending on the formulation, higher-order derivatives via automatic differentiation, which can lead to substantial computational overhead and memory consumption, especially for time-dependent and multi-dimensional problems \cite{raissi2017pinns}. Second, the optimization landscape is often ill-conditioned and PINNs may exhibit slow convergence and sensitivity to loss balancing between data, boundary conditions, and residual terms, and they can fail to resolve sharp gradients, boundary layers, or transport-dominated features without careful sampling and problem-specific stabilization \cite{wang2020understanding,krishnapriyan2021failure}. These difficulties become more pronounced for stiff dynamics or multiscale solutions, where naive collocation strategies and standard network parameterizations may underfit high-frequency components and yield overly smooth predictions \cite{krishnapriyan2021failure}. Consequently, while PINNs offer a flexible, mesh-free framework, their reliability and training cost can be limiting when accurate surrogate evaluations must be produced repeatedly over a broad parameter range.

Given that the present work targets a data-rich setting with large sets of high-fidelity numerical solutions generated by a validated solver, an alternative and often more effective strategy is supervised, field-to-field surrogate modeling. In this approach, the neural network is trained to approximate the input–output map 
\begin{equation*}
    (U_0, \mathbf{c}) \mapsto U_T
\end{equation*}
directly from simulation pairs, avoiding repeated residual evaluation during training and enabling highly efficient inference. Encoder–decoder convolutional architectures with skip connections, such as U-Net \cite{ronneberger2015unet}, are particularly well suited to this task because they represent solution fields on fixed grids and capture multiscale spatial structure through hierarchical feature extraction. U-Net-type models have therefore become common baselines for surrogate modeling of PDE-governed dynamics on structured domains, including in fluid and transport problems \cite{ribeiro2021deepcfd,le2021unet,taccari2022unet}. In the present setting, this choice is further motivated by the need to generalize across a family of PDE instances via a conditioning mechanisms such as Feature-wise Linear Modulation (FiLM) \cite{perez2018film} allowing the network to adapt intermediate representations as a function of the parameter vector $\mathbf{c}$, while explicit coordinate encoding can mitigate translation-invariance biases of pure convolutions \cite{liu2018coordconv}.

A further competitive paradigm is neural operator learning, which seeks to approximate the solution operator mapping between function spaces rather than a single discretization-dependent map. Representative approaches include DeepONet \cite{lu2021deeponet} and Fourier Neural Operators (FNO) \cite{li2021fno}, which have shown strong performance and can generalize across discretizations and resolutions. Neural operators are particularly attractive when one aims to learn resolution-invariant mappings or to transfer across meshes and geometries. However, they typically introduce additional architectural and training complexity (e.g., spectral convolutions, operator-specific parameterizations) and often require careful tuning to achieve consistent gains over strong convolutional baselines in fixed-grid settings. In contrast, for the present objective we required fast prediction at a fixed resolution and fixed time horizon over a prescribed, low-dimensional coefficient family. Thus, a conditioned U-Net provides a favorable trade-off between expressivity, computational efficiency, and implementation simplicity, while remaining compatible with automatic differentiation for downstream sensitivity-based parameter inference.

Our methodological contribution is the incorporation of explicit parameter conditioning and spatial coordinate information into the U-Net surrogate and the application to complex, biologically motivated, nonlinear PDE-governed processes. Parameter conditioning is achieved via the Feature-wise Linear Modulation (FiLM) introduced by Perez et al. \cite{perez2018film}, which introduces affine transformations of the intermediate feature maps based one the conditioning vector $\mathbf{c}$. This design enables a single network to represent an entire family of convection-diffusion-reaction dynamics across the prescribed coefficient range, rather than training separate surrogates for each parameter setting. Moreover, the inclusion of spatial coordinate channels via CoordConv introduced by Liu et al. \cite{liu2018coordconv} improves the networks ability to represent location-dependent effects that are not translation-invariant, consistent with known limitations of standard convolutional layers. Further, we include a residual block structure as base blocks to the U-Net architecture, which allows the learning algorithm to adaptively learn characteristics and features on different resolution scales via a skip-ahead connection. Together, this yields a parameter-conditioned U-Net surrogate that supports accurate predictions over a range of PDE coefficients and facilitates sensitivity-based analysis with respect to the governing parameter functions.

\medskip

The paper is structured as follows. In Section \ref{sec:pde} we give an overview on the considered PDE model including the coefficient functions and initial conditions,  suitable parametrized in order to enable their later randomization. In Section \ref{subsec:discr} we present the numerical discretization scheme including second order approximation of boundary conditions and the application of the Runge-Kutta-Fehlenberg method with time-step control. In Section \ref{subsec:convstudy} the spatial error of the implemented numerical solver is validated demonstrating the expected second order convergence rate. 
The strategy for generating training and test data to develop our surrogate model is described in Section \ref{subsec:Trainingdata}. 
Section \ref{sec:DNN} introduces the DNN surrogate with the architectural components in Section \ref{sec:architectur} and the training details in Section \ref{sec:training} to allow easy reproducibility. This is followed by the training results (Section \ref{sec:unetresults}) and the investigation into the trained surrogates generalization performance based on the neural networks input mechanisms. We discuss and conclude the papers results in Section \ref{sec:conclusion} and propose follow-up work in the outlook Section \ref{sec:outlook}.

\section{PDE Model and Numerical Solver} \label{sec:pde}
To model the dynamics of phenotypic change of malignant cells in two dimensional phenotypic space, 
we consider following nonlinear partial differential equation of convection-diffusion-reaction type 
describing the population density function $u:[0,T]\times\overline{\Omega}\rightarrow \mathbb{R} \text{ with } T \in \mathbb{R}^{+}$ on a rectangular domain $\Omega =(0,L_x)\times(0,L_y) \subseteq \mathbb{R}^2$,

\begin{eqnarray} \label{eqn:model} 
    \frac{\partial}{\partial t} u & =& \nabla (D \nabla u -v \cdot u) + f(u) \quad \text{ in } \Omega \\ 
	\frac{\partial}{\partial n} u & = &0 \qquad \text{ on } \partial \Omega \nonumber
\end{eqnarray}

with the following model parameter functions: 
\begin{itemize}
    \item diffusion matrix $D:\Omega\rightarrow \mathbb{R}^{2\times2}$, where $D(x,y)$ is symmetric and positiv definit on $\Omega$, \\
    \begin{equation*}
        D(x,y)=\begin{bmatrix}
		      d^{1,1}(x,y) & d^{1,2}(x,y) \\
		      d^{1,2}(x,y) & d^{2,2}(x,y) 		
	    \end{bmatrix},
    \end{equation*}
    \item convection velocity $v:\Omega\rightarrow \mathbb{R}^{2}$,
    \begin{equation*}
        v(x,y)=\begin{bmatrix}
		      v^1(x,y) \\
		      v^2(x,y)	
	    \end{bmatrix},
    \end{equation*}
    \item nonlinear reaction term $f:\mathbb{R} \rightarrow \mathbb{R}, \, f(u)=f_0 \cdot u\cdot(f_1-u) \text{ with } f_0,f_1 \in \mathbb{R}^{+}$.
\end{itemize}

The choice of the model is inspired by the work of Chrisholm et al. \cite{chrisholm} on modeling phenotypic 
cell plasticity in a two dimensional phenotypic space structured by proliferation and survival potential. Here the biological interpretation of the nonlinear reaction term $f$ includes proliferation and death of cells; the diffusion term corresponds to non genetic instability and the convection term describes the stress-induced adaptation of the cells.\\
Another similar convection-diffusion-reaction model in one phenotypic dimension of drug resistance can be found in \cite{browning}. A simpler diffusion-reaction mode in one phenotypic dimension (spanning between high proliferation rate to high drug resistance level) can be found in \cite{almeida}. \\
Tables \ref{tab:parameters} and \ref{tab:parameterfunctions} 
provide an overview of the specific model parameter functions used in this study. 
Note, that later on, some parameters of functions 
$D,v,f$ are randomized in order to generate a comprehensive set of numerical training data for the parameter conditioned DNN model, cf. Table \ref{tab:parameterfunctions}.\\

\subsection{Initial Conditions}
\label{sec:initialcondtion}
\begin{figure}[htbp]
	\centering
	\begin{subfigure}[b]{0.45\textwidth}
		\centering
		\includegraphics[width=\textwidth]{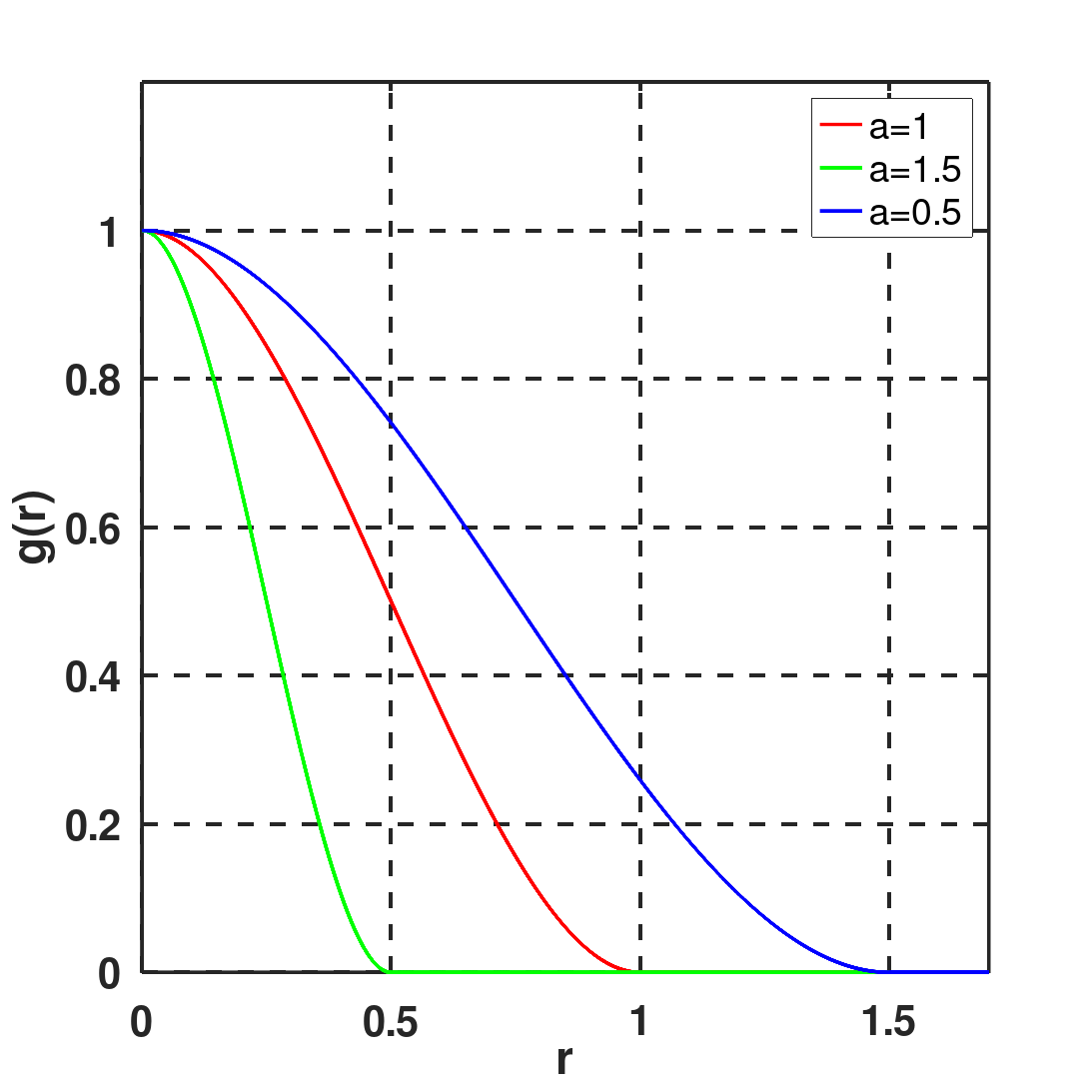}
		\caption{Graph of $g$ for fixed height $h:=H=1$}
		\label{fig:plot1}
	\end{subfigure}
	\hfill
	\begin{subfigure}[b]{0.45\textwidth}
		\centering
		\includegraphics[width=\textwidth]{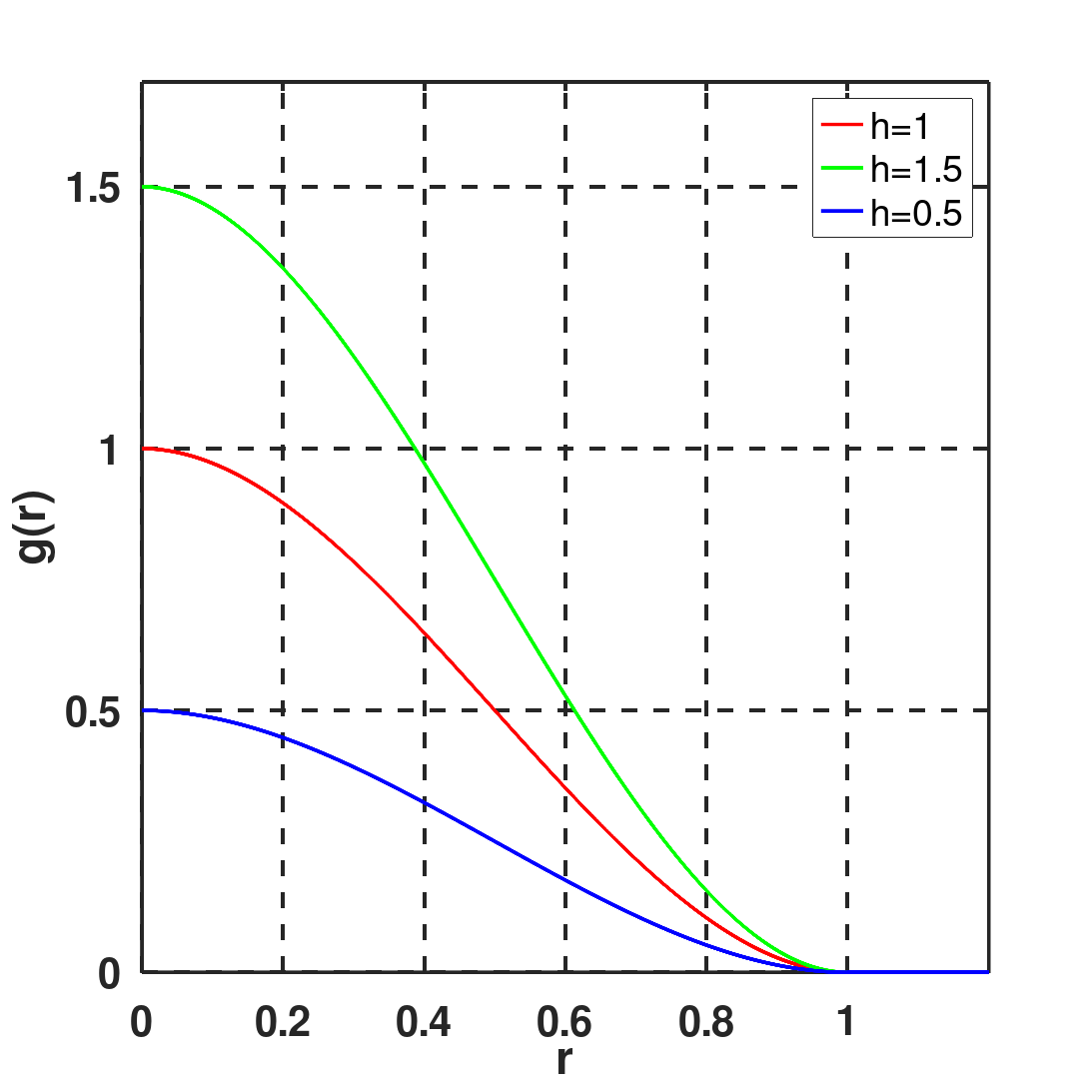}
		\caption{Graph of $g$ for fixed with parameter $a=1$}
		\label{fig:plot2}
	\end{subfigure}
	\caption{Graphical representation of function $g_{H,a}$ for varying parameters $H,a$.}
	\label{fig:two_plots}
\end{figure}

In order to create randomized initial conditions, we consider the initial distribution $u(t=0,\cdot,\cdot)$ to be a sum of several hill-like functions. We construct them via a function $g:\mathbb{R}^+_0\rightarrow \mathbb{R}^+_0$ of the following form
\begin{equation*}
    g(r) = g_{H,a}(r) = \left\{ 
		\begin{array}{ccr} 
		    \frac{2\cdot H}{a^3}\cdot (r-a)^2\cdot (r+0.5\cdot a) &\text{ for }& 0 \leq r < a \\ 
		    0 &\text{ for }& r \geq a
		\end{array}
        \right.
\end{equation*}
where $ H, a \in \mathbb{R}^+$, see Fig. \ref{fig:two_plots} showing $g$ for different parameter values of $H$ and $a$. \\

We now set $G:\mathbb{R}^2\rightarrow \mathbb{R}^+_0$ as
\begin{equation*}
    G(x,y)=G_{H,a,x_{\max},y_{\max}}(x,y) = g_{H,a}\left (
    \norm{\begin{pmatrix} x \\ y \end{pmatrix}
    - \begin{pmatrix} x_{\max} \\ y_{\max} \end{pmatrix}}
    \right ).
\end{equation*}

Note that $ H \in \mathbb{R}^+$ scales the height of the hill function $G$, whereas $a\in\mathbb{R}^+$ defines its width and $(x_{\max},y_{\max})$ are the coordinates of the hill center, i.e., of the maximum point.
Moreover, $G$ is continuously differentiable w.r.t to $x, y$ and has bounded support $supp(G_{H,a,x_{\max},y_{\max}})=B_a(x_{\max},y_{\max})$.

We now set $u(t=0,x,y)$ as the sum of several hill-type functions
\begin{equation} \label{eqn:initial}
    u(t=0,x,y)=\sum_{i=1}^n G_{H_i,a_i,x_{\max}^i,y_{\max}^i} (x,y)
\end{equation}
and sample
\begin{itemize}
	\item[$i)$] the number $n$ of hill-function summands,  where $n\in \{5,6,...,15\}$
	\item[$ii)$] the height $H_i$ of every hill function, where $H_i\in (0,1)$
	\item[$iii)$] the position $(x_{\max}^i,y_{\max}^i)$ of the maximum $(x_{\max}^i,y_{\max}^i)\in [\frac{1}{5}L_x,\frac{4}{5}L_x]\times[\frac{1}{5}L_y,\frac{4}{5}L_y]$
	\item[$iv)$] the size of the support of hill-functions by choosing
    \begin{equation} \label{eqn:interval_ai}
        a_i \in \left [ \frac{1}{10}\min(L_x,L_y),\min(x_{\max}^i,L_x-x_{\max}^i,y_{\max}^i,L_y-y_{\max}^i) \right ].
    \end{equation}
    Note that:
    \begin{itemize}
        \item $\left [ \frac{1}{10}\min(L_x,L_y),\min(x_{\max}^i,L_x-x_{\max}^i,y_{\max}^i,L_y-y_{\max}^i) \right ] \neq \emptyset$\\ due to the position of 
        $(x_{\max}^i,y_{\max}^i)$ within our rectangle domain, see $iii)$ above
        \item $a_i < \min(x_{\max}^i,L_x-x_{\max}^i,y_{\max}^i,L_y-y_{\max}^i)$ ensures $supp(G) \subset\overline{\Omega}$,  i.e.,
        \\the zero-Neumann boundary condition at $t=0$ is satisfied 
        \item $a_i > \frac{1}{10}\min(L_x,L_y)$ ensures the derivatives of $G$ to be bounded for every choice of $a_i$ satisfying Equation (\ref{eqn:interval_ai})
    \end{itemize}
    In practice, we construct $a_i$ by sampling 
    $R_i \in [0,1]$ and set \\
    \begin{equation}\label{eqn:ai}
        a_i=\frac{1}{10}\min(L_x,L_y) +R_i\cdot \left (\min(x_{\max}^i,L_x-x_{\max}^i,y_{\max}^i,L_y-y_{\max}^i)-\frac{1}{10}\min(L_x,L_y) \right )
    \end{equation}
\end{itemize}

\begin{figure}[H]
	\centering
	\includegraphics[scale=0.55]{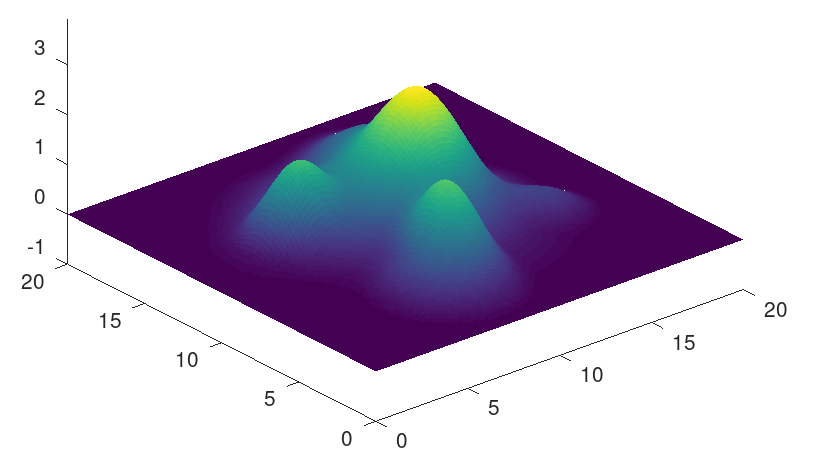}
	\caption{Initial condition (\ref{eqn:initial}) used in the the convergence study with randomized parameters from Table \ref{tab:init}}
	\label{fig:init}
\end{figure}

An exemplary initial function with randomly sampled parameters given in Table \ref{tab:init} is depicted in Figure \ref{fig:init}.

\subsection{Numerical Discretization} \label{subsec:discr}
To solve our PDE numerically, we first rewrite Equation $(1)$ using the chain rule as
\begin{align}
    \begin{array}{ccl}
        \frac{\partial}{\partial t} u & = &
        -(\partial_xv^1+\partial_yv^2)u
        +(\partial_xd^{11}+\partial_yd^{12}-v^1) \partial_xu
        +(\partial_xd^{12}+\partial_yd^{22}-v^2) \partial_yu \\
        & &
        +d^{11}\partial_{xx}u 
        + 2 \cdot d^{12}\partial_{xy}u
        + d^{22}\partial_{yy}u + f(u)
    \end{array}
	\label{eqn:chainrule}
\end{align}
Afterwards, we semidiscretize our PDE by first introducing a uniform grid of size $N\times N$ with grid points $(x_i,y_j)$ for $i,j\in\{0,...,N-1\}$ on our rectangular domain $\Omega$ 
Next, we apply the finite-difference method by using central differences on the interior node and second order forward / backwards stencils on the boundary nodes, see \cite{fornberg}.
By this procedure, we obtain the following equations for the numerical semi-discrete approximation $\overline{u}_{i,j}(t) \approx u(t,x_i,y_j)$ \\

For the interior nodes with $i,j\in\{1,...,N-2\}$ it holds
\begin{align}
    \begin{array}{ccl}
          \overline{u}\,'_{i,j}(t) & = &
        -(\mathscr{D}_x(v^1_{i,j})+\mathscr{D}_y(v^2_{i,j}))\overline{u}_{i,j}(t) \\
        & & +(\mathscr{D}_{x}(d^{11}_{i,j}) + \mathscr{D}_{y}(d^{12}_{i,j}) - v^1_{i,j}) \mathscr{D}_x(\overline{u}_{i,j}(t)) \\ 
        & & +(\mathscr{D}_x(d^{12}_{i,j}) + \mathscr{D}_y(d^{22}_{i,j}) - v^2_{i,j}) \mathscr{D}_y(\overline{u}_{i,j}(t)) \\
        & & +d^{11}_{i,j} \cdot \mathscr{D}_{xx} (\overline{u}_{i,j}(t)) + 2d^{12}_{i,j} \cdot \mathscr{D}_{xy} (\overline{u}_{i,j}(t)) + d^{22}_{i,j} \cdot \mathscr{D}_{yy} (\overline{u}_{i,j}(t)) \\
        & = : & F_{i,j}(\overline{u}_{i,j}(t))
    \end{array}
    \label{eqn:semidiscretized}
\end{align}
where we use the abbreviations $d^{11}_{i,j} = d^{11}(x_i,y_j)$, $d^{12}_{i,j} = d^{12}(x_i,y_j)$, $d^{22}_{i,j} = d^{22}(x_i,y_j)$, $v^{1}_{i,j} = v^{1}(x_i,y_j)$, $v^{2}_{i,j} = v^{2}(x_i,y_j)$ and $\mathscr{D}$ denote the corresponding finite-difference operators
\begin{eqnarray*}
    \mathscr{D}_{x}(q_{i,j}) &:=& \frac{1}{2\cdot \Delta x}(q_{i+1,j}-q_{i-1,j}),\\
    \mathscr{D}_{y}(q_{i,j}) &:=& \frac{1}{2\cdot \Delta y}(q_{i,j+1}-q_{i,j-1}), \\
    \mathscr{D}_{xx}(q_{i,j}) &:=& \frac{1}{\Delta x^2} (q_{i+1,j}-2\cdot q_{i,j}+q_{i-1,j}), \\
    \mathscr{D}_{yy}(q_{i,j}) &:=& \frac{1}{\Delta y^2}(q_{i,j+1}-2\cdot q_{i,j}+q_{i,j-1}),
    \\
    \mathscr{D}_{xy}(q_{i,j}) &:=& \frac{1}{4 \cdot \Delta x \cdot \Delta y} (q_{i+1,j+1}-q_{i+1,j-1}-q_{i-1,j+1}+q_{i-1,j-1}). \\
\end{eqnarray*}

For the boundary nodes with $i\in \{1,...,N-2\}$ (corners excluded on the top and bottom edge) the Neumann boundary condition is approximated by second order  scheme,
\begin{align}
    \begin{array}{rcc}
        \frac{1}{\Delta y} (-1.5\cdot \overline{u}_{i,0}(t)+2\cdot \overline{u}_{i,1}(t)-0.5 \cdot \overline{u}_{i,2}(t)) & = & 0, 
        \vspace{0.2cm} \\
        \frac{1}{\Delta y} (1.5\cdot \overline{u}_{i,N-1}(t)-2\cdot \overline{u}_{i,N-2}(t)+0.5 \cdot \overline{u}_{i,N-3}(t)) & = & 0.
    \end{array}
    \label{eqn:boundaryone}
\end{align}

Analogously, for the boundary nodes with $j\in \{0,...,N-1\}$ (corners included on the left and right edge) it holds
\begin{align}
    \begin{array}{rcc}
        \frac{1}{\Delta x} (-1.5\cdot \overline{u}_{0,j}(t)+2\cdot \overline{u}_{1,j}(t)-0.5 \cdot \overline{u}_{2,j}(t)) & = & 0, 
        \vspace{0.2cm} \\
        \frac{1}{\Delta x} (1.5\cdot \overline{u}_{N-1,j}(t)-2\cdot \overline{u}_{N-2,j}(t)+0.5 \cdot \overline{u}_{N-3,j}(t)) & = & 0.
    \end{array}
    \label{eqn:boundartwo}
\end{align}

Finally,  the  resulting ODE system 
(\ref{eqn:semidiscretized}) 
is  time-discretized by applying the following Runge-Kutta (Fehlenberg) pair of second and third order enabling the step-size control, see \cite{fehlberg},
\begin{table}[htbp]
	\centering
	\begin{tabular}{c|c c c}
		0 &  &  &  \\
		1 & 1 &  &  \\
		$1/2$ & $1/4$ & $1/4$ &  \\
		\hline
		 & $1/2$ & $1/2$ &  \\
		 & $1/6$ & $1/6$ & $2/3$ 
	\end{tabular}
	\caption{Butcher tableau of the used Runge-Kutta-Scheme}
    \label{tab:butcher}
\end{table} \\
Denoting the fully discretized approximation of the solution at time $t_m$ with $m\in \{1,...,M\}$ by $U_{i,j,m}$, the Fehlenberg method yields the following approximation formulas for the numerical solution on the interior nodes $(i,j)$ with $i,j \in \{1,...,N-2\}$:
\begin{align}
    \begin{array}{ccl}
        U_{i,j,m+1} & =& U_{i,j,m}+\Delta t \cdot  (\frac1{6}  k_{1,i,j}+\frac1{6}  k_{2,i,j}+\frac{2}{3}  k_{3,i,j}),\\
        k_{1,i,j} & = & F_{i,j}(U_{i,j,m}) \\
        k_{2,i,j} & = & F_{i,j}(U_{i,j,m}+\Delta t  \cdot k_{1,i,j}), \\
        k_{3,i,j} & = & F_{i,j}(U_{i,j,m}+\Delta t \cdot (\frac1{4}  k_{1,i,j}+ \frac1{4}  k_{2,i,j})).
    \end{array}
    \label{eqn:fullydiscretized}
\end{align}

Note that due to the spatial discretization of derivatives,  for the evaluation of the right hand side term $F_{i,j}$ of the ODE-system (\ref{eqn:semidiscretized})
the values of  $U_m, \, k_1$ and $k_2$ 
in all to $i,j$- adjacent nodes $(i\pm 1,j\pm 1), (i,j\pm 1), (i\pm 1,j)$ are necessary 
for calculating the Runge-Kutta stages $k_{2,i,j}$ and $k_{3,i,j}$ 
in  (\ref{eqn:fullydiscretized}).
More precisely,  the arguments $U_{i,j,m}+\Delta t \cdot k_{1,i,j}$ and $U_{i,j,m}+\Delta t \cdot (\frac1{4}  k_{1,i,j}+ \frac1{4}  k_{2,i,j})$ of $F_{i,j}$  need to be evaluated on all grid nodes with $i,j \in \{0,...,N-1\}$ including the boundary nodes, 
and thus the evaluation of function $F_{i,j}$ requires special attention, see e.g., \cite{pathria}.
We  circumvent this obstacle by interpreting these
arguments 
as predicted intermediate values of the numerical approximation at intermediate times, i.e., $\overline{u}_{i,j}(t_m + \Delta t)\approx U_{i,j,m}+\Delta t \cdot k_{1,i,j}$,   $\overline{u}_{i,j}(t_m + \frac1{2} \cdot \Delta t)\approx U_{i,j,m}+\Delta t \cdot (\frac1{4} \cdot k_{1,i,j}+ \frac1{4}\cdot k_{2,i,j})$, 
cf. the coefficients of the applied Runge-Kutta scheme in Table \ref{tab:butcher}. This gives the opportunity to impose the (discretized) boundary conditions (\ref{eqn:boundaryone}) and (\ref{eqn:boundartwo}) directly for $U_{i,j,m}+\Delta t \cdot k_{1,i,j}$ and $U_{i,j,m}+\Delta t \cdot (1\frac1{4} \cdot k_{1,i,j}+ \frac1{4} \cdot k_{2,i,j})$, to extend these intermediate values from the adjacent interior to the boundary nodes.\\

\subsubsection*{Time-step control}
The calculated solution values $U_{i,j,m+1}$ (of third order) and the second order approximation $U^*_{i,j,m+1} = U_{i,j,m}+\Delta t  (\frac{1}{2}  k_{1,i,j}+\frac1{2} k_{2,i,j})$ are used 
for step-size control as follows: In  each time step, the solution is computed according to (\ref{eqn:fullydiscretized}) at first with some initial step-size 
$\Delta t=\Delta t_{init}$. Afterwards an optimal time-step $\Delta t_{opt}$ is predicted based on $\Delta t_{init}$, on temporal error estimation  by $\| U^*_{m+1}-U_{m+1}\|_\infty$  and on a chosen error tolerance $tol$  via the following, 
\begin{equation}
    \Delta t_{opt} = \left\{
    \begin{array}{ll}
         0.9 \cdot \Delta t_{init} \cdot 2
         & \mbox{if} \  \, \left ( \frac{tol}{\|U^*_{m+1}-U_{m+1}\|_\infty} \right )^\frac13>2 \\
         0.9 \cdot \Delta t_{init} \cdot 0.3
         & \mbox{if} \  \, \left ( \frac{tol}{\|U^*_{m+1}-U_{m+1}\|_\infty} \right )^\frac13<0.3 \\
         0.9 \cdot \Delta t_{init} \cdot \left ( \frac{tol}{\|U^*_{m+1}-U_{m+1}\|_\infty} \right )^\frac13
         & \mbox{else,}  \\
    \end{array}
    \right.
    \label{eqn:topt}
\end{equation} 
see e.g., in \cite{strehmel}. 
If the predicted step-size $\Delta t_{opt}\geq \Delta t_{init}$, the calculation of $U_{m+1}$ is accepted and $\Delta t_{opt}$ is used in  the next time step as  initial guess for $\Delta t_{init}$. Otherwise
$U_{m+1}$ is discarded and $U_{m+1},U^*_{m+1}$ are recalculated  according to (\ref{eqn:fullydiscretized}) using $\Delta t_{opt}$. Our step-size control strategy enables certain, solution-dependent
 step-size variation within a defined range, while suppressing a strong variation of the time-step. 
In our implementation  $tol=10^{-6}$ is chosen,  guaranteeing a similar range for the local temporal discretization error.

\subsection{Mesh Convergence Study}\label{subsec:convstudy}

The above described discretization method was implemented in C++. To validate the accuracy of our computational model and its implementation, we experimentally analyse the behaviour of the numerical discretization error with regard to mesh refinement and validate the experimental order of convergence (EOC).

To achieve this, we consider an exemplary initial condition, chosen as a sum of $15$ hill-like functions $G$ with randomly sampled values for $H$, $x_{\max}$, $y_{\max}$ and $R$ presented in the Table \ref{tab:init}. 
\begin{table}[H]
	\centering
	\begin{tabular}{c|c|c|c|c}
		$i$ & $H_i$ & $x_m^i$ & $y_m^i$ & $R_i$ \\
		\hline 
		1 & 0.45 & 10.18 & 9.28 & 0.81 \\
		2 & 0.21 & 5.84 & 11.39 & 0.00 \\
		3 & 0.73 & 9.01 & 12.17 & 0.84 \\
		4 & 0.83 & 4.98 & 11.55 & 0.21 \\
		5 & 0.39 & 15.37 & 7.23 & 0.28 \\
		6 & 0.78 & 7.39 & 4.14 & 0.98 \\
		7 & 0.82 & 8.78 & 6.12 & 0.16 \\
		8 & 0.26 & 5.22 & 11.61 & 0.79 \\
		9 & 0.75 & 11.60 & 11.18 & 0.32 \\
		10 & 0.12 & 10.31 & 11.06 & 0.70 \\
		11 & 0.16 & 14.32 & 10.68 & 0.32 \\
		12 & 0.53 & 11.06 & 9.17 & 0.37 \\
		13 & 0.45 & 8.65 & 4.94 & 0.25 \\
		14 & 0.68 & 12.74 & 15.33 & 0.94 \\
		15 & 0.46 & 11.93 & 11.26 & 0.15 
	\end{tabular}
	\caption{Parameters of the hill-like functions $G$ of the initial condition used in the error convergence study}
	\label{tab:init}
\end{table} 

\vspace{-5pt}
\begin{figure}[H]
	\centering
	\includegraphics[scale=0.20]{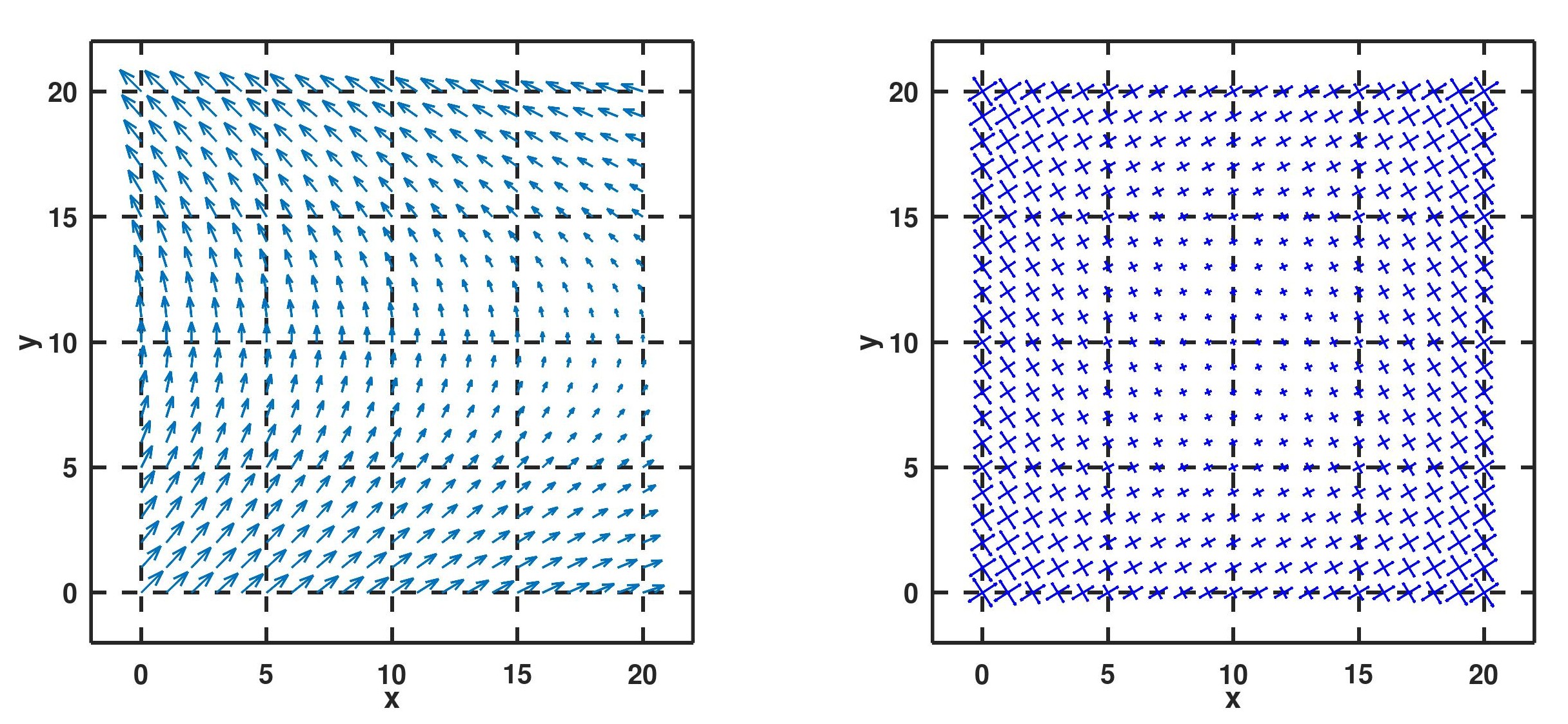}
	\caption{Visual representation of the velocity function $v$ (on the left) and the eigenvectors of the diffusion matrix $D$ scaled by their respective eigenvalues (on the right), used in the convergence study.}
	\label{fig:parameter}
\end{figure}

The parameter functions 
of our model (\ref{eqn:model}) are exemplary chosen as a paraboloid-type diffusion matrix function $D$ and a bi-linear transport velocity function $v$. They are specified in Table \ref{tab:parameters} and graphically represented in Figure \ref{fig:parameter}.

\begin{table}[htbp]
	\centering
	\begin{tabular}{c|l|c}
		\bf Parameter & \bf Description & \text{\bf Numerical value / Function formula} \\
		\hline 
		$L_x=L_y:=L $ & length of computational domain & $20$ \\
	    \hline
		$T$ & final time & 1.5 \\
		\hline
		$v_1(x,y)$ & components of advective velocity & $3-6\cdot \frac{y}{L_y}$ \\
		$v_2(x,y)$ & & $3-2\cdot \frac{x}{L_x}$ \\
        \hline
        $d_{1,1}(x,y)$ & components of diffusion matrix &  $\frac{1}{2} + 4(\frac{x}{L_x}-\frac{1}{2})^2 + 4(\frac{y}{L_y}-\frac{1}{2})^2$ \\
		$d_{1,2}(x,y)$ & & $\frac{1}{10} + (\frac{x}{L_x}-\frac{1}{2})^2 + 2(\frac{y}{L_y}-\frac{1}{2})^2$ \\
		$d_{2,2}(x,y)$ & & $\frac{1}{4} + 4(\frac{x}{L_x}-\frac{1}{2})^2 + 2(\frac{y}{L_y}-\frac{1}{2})^2$ \\
		\hline
		$f(u)$ & reaction term & $\frac{1}{2} \cdot u \cdot (2-u)$ \\
	\end{tabular}
	\caption{Table of the used model parameters in Equation (\ref{eqn:model}) }
	\label{tab:parameters}
\end{table} 

In the following we denote the numerical approximation of the solution on an $N \times N$-grid with stepsize $h = \Delta x = \Delta y$ at final time $T$ and point $(x_i,y_j)$ as $U^{h}_{i,j,M}$. 
Then we define\\
\begin{align}
    \begin{array}{clcc}
        & \left \|U^{h}-U^{h/2} \right \|_2 & = &
        \sqrt{
        \frac{1}{N^2}\sum_{i,j=0}^{N-1} |U^{h}_{i,j,M}-U^{h/2}_{2i,2j,M}|^2
        } \\
        & \left \|U^{h}-U^{h/2} \right \|_\infty & = & 
        \max_{0 \leq i,j \leq N-1}\ |U^{h}_{i,j,M}-U^{h/2}_{2i,2j,M}|
    \end{array}
\end{align}
and evaluate the experimental order of convergence as follows, see \cite{hundertmark}:
\begin{equation}
	EOC(h_i)=\log_2 \left ( \frac{\left \|U^{h}-U^{h/2} \right \|}{\left \|U^{h/2}-U^{h/4} \right \|} \right )
	\label{eqn:EOC}
\end{equation}

A sequence on five meshes has been constructed by regular refining the original mesh of $51 \times 51 $ grid points by halving the step sizes, starting from mesh size $h=0.4 \ (h/L=0.02)$ with the final mesh size $h=0.025 \ (h/L=0.00125)$. 

The following table shows the absolute error of the numerical solution w.r.t. the refined one and the experimental order of convergence using $\| \cdot \|_2$ and $\| \cdot \|_\infty$

\begin{table}[H]
	\centering
	\begin{tabular}{c||c|c|c|c}
		$h/L$ & $\left \|U^{h}-U^{h_/2} \right \|_{L^2}$ 
		& $\left \|U^{h}-U^{h/2} \right \|_{L^\infty}$ & $ \text{EOC in }\left \|. \right \|_{L^2}$ & \text{EOC in }$\left \|. \right \|_{L^\infty}$  \\[2pt]
		\hline \hline
		0.02 & 0.00635 &  0.03228 & 2.0783 & 2.1168 \\[10pt]
		0.01 & 0.00150 &  0.00744 & 2.0734 & 2.1094 \\[10pt]
		0.005 & 0.00036 &  0.00172 & 2.0580 & 2.0901 \\[10pt]
		0.0025 & 0.00009 & 0.00041 & / & / \\[10pt]
	\end{tabular}
	\caption{Numerical error of the solutions on two subsequent meshes and the experimental order of convergence confirm the theoretical second order convergence rate.}
    \label{tab:L2erros}
\end{table} 

\begin{figure}[H]
	\centering
	\includegraphics[width=0.5\textwidth]{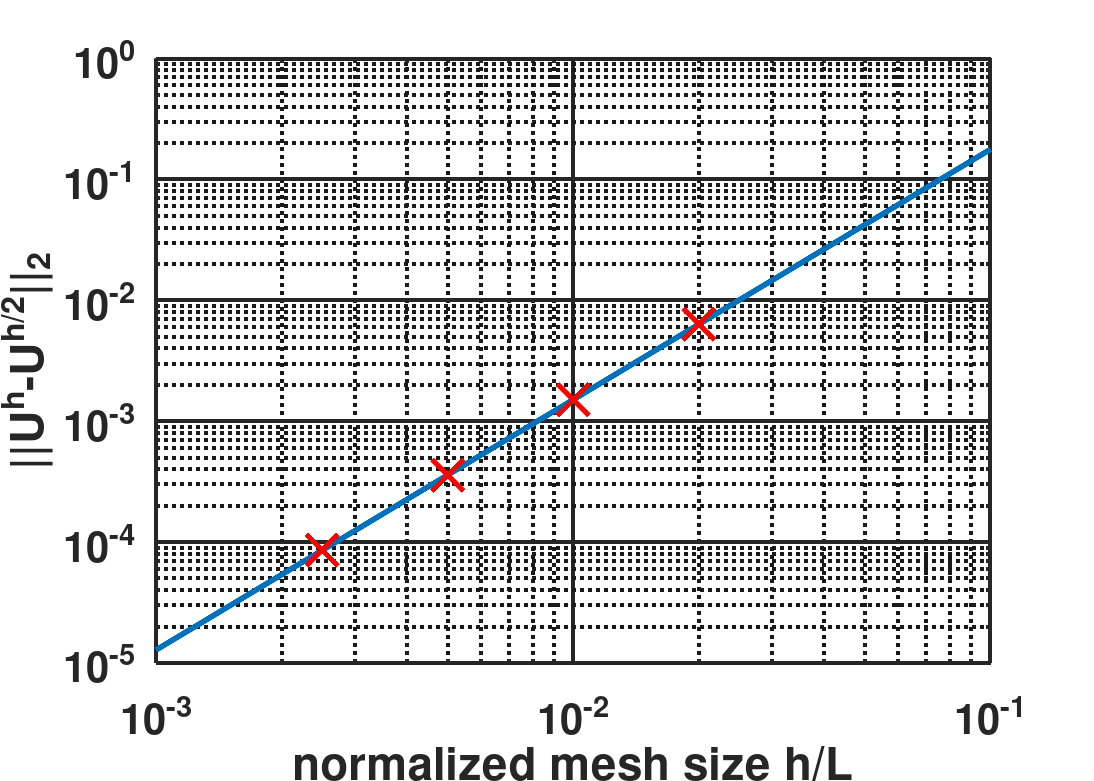}
	\caption{The decay of the $L^2$-errors of our discretization scheme for different mesh resolutions (red crosses) indicates the second order decrease (blue line). }
    \label{fig:L2erros}
\end{figure}

The following table shows the relative error with respect to the reference solution $U_{ref}=U^{0.025}$ along with the average timestep while computing and the computational time to generate the solution.
\begin{table}[H]
	\centering
	\begin{tabular}{c||c|c|c|c}
		$h/L$ & $\frac{\left \|U^{h_i}-U_{ref} \right \|_{L^2}}{\left \|U_{ref} \right \|_{L^2}}$ & $\frac{\left \|U^{h_i}-U_{ref} \right \|_{L^\infty}}{\left \|U_{ref} \right \|_{L^\infty}}$ & avg $\Delta t$ & comput. time \\[10pt]
		\hline \hline
		0.02 & 0.00949 & 0.0213 & 0.00688 & 0.26 s. \\[10pt]
		0.01 & 0.00223 & 0.0049 & 0.00625 & 1.38 s.\\[10pt]
		0.005 & 0.00050 & 0.0011 & 0.00168 & 18 s.\\[10pt]
		0.0025 & 0.00010 & 0.0002 & 0.00040 & 192 s.\\[10pt]
		\hline \hline
		0.00125 & 0 & 0 & 0.00010 & 3022 s. \\[2pt]
	\end{tabular}
	\caption{The decay of the relative discretization error, the used average time step size and the computational time in seconds.}
    \label{tab:rel_errors}
\end{table} 

\subsection{Computation of Training and Test Data}\label{subsec:Trainingdata}

For the set up of the DNN surrogate model described in Section \ref{sec:DNN} a set of training and test data was generated as follows.
The training set consists of 10000 pairs of randomly sampled initial conditions $U_{0}^{(k)}$ ($k=1,...,10000$) and their respective final solutions $U_{M}^{(k)}$ at the final timestep for randomly sampled model parameter functions. The solutions were generated with our second order numerical solver on a mesh of $2^8 \times 2^8 = 256 \times 256$ grid points. The choice of the grid size as power of $2$ was purposefully. It allows down sampling and upscaling operations in the DNN to produce same size matrices in the encoder and decoder paths, thus allowing skip connections without the necessity for any shape manipulations.

This grid size corresponds to $h/L \approx 0.0039$ lying
between the mesh-size of third and fourth grid of our convergence study. The numerical relative $L^2$ error (w.r.t. the highly resolved reference 
solution) can be therefore expected below $0.05\, \%$, and the relative $L^\infty$ error below $0.1\, \%$
cf. Table \ref{tab:rel_errors}.
The initial conditions $U_{0}^{(k)}$ were chosen by independent uniform sampling using Equation (\ref{eqn:initial}) for randomized combination of hill-like functions inside of the constraints described in Subsection \ref{sec:initialcondtion}. For each of the $10000$ initial solutions a different set of model parameter functions $D,v,f$ was randomly chosen in the form of a parameterized family of functions shown in Table \ref{tab:parameterfunctions}, where the individual parameters $\bm{c}^{(k)}= (c^{(k)}_1,c^{(k)}_2,c^{(k)}_3,c^{(k)}_4)$ were uniformly sampled with $c^{(k)}_1 \in [0,6]$, $c^{(k)}_2 \in [-1,3]$, $c^{(k)}_3 \in [1,9]$, $c^{(k)}_4 \in [1,3]$. 

For the test set, another $2500$ pairs of randomized initial conditions and their respective final solution for randomized model parameter functions were generated with the numerical solver on the same $256 \times 256$ grid. 
To generate these pairs, $50$ different randomly sampled initial conditions $U_0^{(k_1)}$ and $50$ different coefficients combinations $ \bm{c}^{(k_2)}=(c_1^{(k_2)},c_2^{(k_2)},c_3^{(k_2)},c_4^{(k_2)})$ were chosen, all from the same distribution as for the training set. 
The final solutions $U_M^{(k_1,k_2)} =U_{M}(U_{0}^{(k_1)}, \bm{c}^{(k_2)})$ were computed for each possible combination of $k_1$-th initial solution and $k_2$-th parameter choice, resulting in new $2500$ test data pairs in total. For computational efficiency we downsample training and test datasets to $64 \times 64$ grids for the training of the DNN surrogate, therefore, the surrogate approximates a coarse-grained operator instead. For simplicity, we will continue to refer to the initial population density and final solution in the $64 \times 64$ resolution as $X_0$ and $X_M$.
\begin{table}[htbp]
	\centering
	\begin{tabular}{c|l|c}
		\bf Parameter function & \bf Description & \text{ \bf Function formula 
        } \\
		\hline 
		$v_1(x,y)$ & advective velocities & $3-c_1\cdot \frac{y}{L_y}$ \\
		$v_2(x,y)$ & & $c_2-2\cdot \frac{x}{L_x}$ \\
		\hline
        $d_{1,1}(x,y)$ & components of & $0.5 \cdot c_3 + 4(\frac{x}{L_x}-\frac{1}{2})^2 + 4(\frac{y}{L_y}-\frac{1}{2})^2$ \\
		$d_{1,2}(x,y)$ & diffusion matrix & $\frac{1}{10} + (\frac{x}{L_x}-\frac{1}{2})^2 + 2(\frac{y}{L_y}-\frac{1}{2})^2$ \\
		$d_{2,2}(x,y)$ & & $\frac{1}{4} + 4(\frac{x}{L_x}-\frac{1}{2})^2 + 2(\frac{y}{L_y}-\frac{1}{2})^2$ \\
		
		\hline
		$f(u)$ & reaction term & $\frac{1}{2} \cdot u \cdot (c_4 -u)$ \\
	\end{tabular}
	\caption{Table of the used PDE-model parameter functions in Equation (\ref{eqn:model}) with randomized coefficients $c_1, c_2,c_3, c_4$.}
	\label{tab:parameterfunctions}
\end{table}

\section{Deep Neural Network Surrogate Model}\label{sec:DNN}

We consider the dataset $\mathcal{D} = \{(X_0^{(i)}, X_M^{(i)}, \bm{c}^{(i)})\}_{i=1}^N$, where each tuple consists of the initial field $X_0 \in \mathbb{R}^{C \times H \times W}$, a solution field $X_M \in \mathbb{R}^{C \times H \times W}$, and a parameter vector $\bm{c} \in \mathbb{R}^d$ characterizing the PDE coefficients. In our specific case $C=1$, $H=64$ and $W=64$, following Section \ref{subsec:Trainingdata}. The intermediate states $\{U_m\}_{m=1}^{M-1}$ are not considered during training. The objective of the surrogate is to learn a parametric mapping $F_\theta$ such that
\begin{equation*}
  F_\theta(X_0, \bm{c}) \approx X_M,
\end{equation*}
where $\theta$ are all trainable parameters (weights and biases). $F_\theta$ approximates the cumulative effect of the convection-diffusion operator of our PDE model prescribed in Equation \ref{eqn:model} over the time interval $[0, M]$, conditioned on the physical parameters $\bm{c}$.

\subsection{Architecture Overview}\label{sec:architectur}

The architecture follows an encoder-decoder structure with skip connections, commonly referred to as U-Net, first introduced by Ronneberger et al.\cite{ronneberger2015unet}. Originally developed for biomedical image segmentation, U-Net architectures have demonstrated success in learning mappings between structured fields, making them well-suited for PDE surrogate modeling~\cite{ronneberger2015unet}. Our implementation augments the base architecture with parameter conditioning via Feature-wise Linear Modulation (FiLM)~\cite{perez2018film} and coordinate encoding~\cite{liu2018coordconv} and replaces the base building blocks with residual blocks. The full architecture is shown in Figure \ref{fig:unet-architecture}. 
For completeness, let $L$ denote the number of resolution levels with channel multipliers $\{m_\ell\}_{\ell=0}^{L-1}$. 

\subsubsection{Activation Function}
Throughout the network, we employ the Leaky Rectified Linear Unit (LeakyReLU) as the nonlinear activation function~\cite{maas2013leakyrelu}
\begin{equation*}
    \sigma(x) = \text{LeakyReLU}(x) = \begin{cases}
        x & \text{if } x \geq 0, \\
        \alpha_{\text{neg}} x & \text{if } x < 0,
    \end{cases}
\end{equation*}
where $\alpha_{\text{neg}} = 0.01$ is the negative slope coefficient. Unlike the standard ReLU, LeakyReLU permits a small gradient when the unit is not active, mitigating the "dying ReLU" problem where neurons become permanently inactive during training~\cite{maas2013leakyrelu}.

\subsubsection{Normalization}
We employ Group Normalization (GroupNorm)~\cite{wu2018groupnorm} as our normalization strategy. For a feature map $h \in \R^{C' \times H' \times W'}$, where $C', H',W'$ are the number of channels, height and width at the applicable layer, the channels are divided into $G$ groups, and normalization is performed within each group
\begin{equation*}
    \text{GroupNorm}(h) = \gamma \odot \frac{h - \mu_i}{\sqrt{\sigma_i^2 + \epsilon}} + \beta,
\end{equation*}
where $\mu_i$ and $\sigma_i^2$ are the mean and variance computed over spatial dimensions and channels within each group $i \in \{1, \ldots, G\}$, and $\gamma, \beta \in \R^{C'}$ are learnable affine parameters.

GroupNorm offers some advantages over Batch Normalization~\cite{ioffe2015batchnorm} in our setting. Its statistics are computed independently of the batch size, providing stable gradient flow even with small batches~\cite{wu2018groupnorm}. 

\subsubsection{Coordinate Encoding}\label{subsec:coordenc}
Following the CoordConv approach~\cite{liu2018coordconv}, the input is augmented with normalized spatial coordinate channels
\begin{equation*}
    \tilde{X}_0 = [X_0; \bm{\xi}; \bm{\eta}] \in \R^{(C+2) \times H \times W},
\end{equation*}
where $\bm{\xi}_{i,j} = \frac{2i}{H-1} - 1$ and $\bm{\eta}_{i,j} = \frac{2j}{W-1} - 1$ for $i \in \{0, \ldots, H-1\}$, $j \in \{0, \ldots, W-1\}$, yielding coordinate grids $\bm{\xi}, \bm{\eta} \in [-1, 1]^{H \times W}$.

Standard convolutional networks exhibit translational equivariance, meaning they can not inherently distinguish spatial positions~\cite{liu2018coordconv}. For PDE solutions, however, spatial positions carry critical information. Boundary conditions impose position-dependent constraints, and the solution structure often varies systematically across the domain. The explicit coordinate channels enable the network to learn position-dependent features essential for capturing boundary effects and spatially varying solution characteristics.

\subsubsection{Conditioning Embedding}
The physical parameter vector $\bm{c} \in \R^d$ encodes the PDE coefficients (e.g., diffusion strength, velocity magnitude). This vector is mapped to a latent embedding via a three-layer multilayer perceptron (MLP)
\begin{equation*}
    \bm{e} = \text{MLP}(\bm{c}) = W_3 \sigma(W_2 \sigma(W_1 \bm{c} + \bm{b}_1) + \bm{b}_2) + \bm{b}_3 \in \R^{d_e},
\end{equation*}
where $W_1 \in \R^{d_h \times d}$, $W_2 \in \R^{d_h \times d_h}$, $W_3 \in \R^{d_e \times d_h}$ are weight matrices with $d_h = \kappa \cdot d_e$ being the hidden dimension determined by a multiplier $\kappa$, and $\sigma(\cdot)$ denotes the Sigmoid Linear Unit activation~\cite{elfwing2018silu}
\begin{equation*}
    \text{SiLU}(x) = x \cdot \text{sigmoid}(x) = \frac{x}{1 + e^{-x}}.
\end{equation*}
The SiLU activation, also known as Swish~\cite{ramachandran2017swish}, provides smooth gradients which is ideal for the conditioning mechanism. 

\subsubsection{Feature-wise Linear Modulation (FiLM)}\label{subsec:film}
The conditioning embedding $\bm{e}$ modulates intermediate feature representations through Feature-wise Linear Modulation (FiLM)~\cite{perez2018film}. Given a feature map $h' \in \R^{C_{\text{out}} \times H' \times W'}$, FiLM applies an affine transformation
\begin{equation*}
    \text{FiLM}(h', \bm{e}) = h' \odot (1 + \bm{\gamma}(\bm{e})) + \bm{\beta}(\bm{e}),
\end{equation*}
where $\bm{\gamma}, \bm{\beta} \in \R^{C_{\text{out}}}$ are computed via a linear projection
\begin{equation*}
    [\bm{\gamma}; \bm{\beta}] = W_{\text{film}} \bm{e} \in \R^{2C_{\text{out}}},
\end{equation*}
and $\odot$ denotes the element-wise multiplication.

The multiplicative term $(1 + \bm{\gamma})$ enables the network to selectively scale feature channels based on the physical parameters, while the additive term $\bm{\beta}$ provides channel-wise biases. This formulation is more expressive than simple additive conditioning, allowing the network to amplify or suppress features depending on the PDE regime~\cite{perez2018film}. The transformation is initialized to identity ($\bm{\gamma} = \bm{0}$, $\bm{\beta} = \bm{0}$) for stable training initialization.

\subsubsection{Residual Block with FiLM Conditioning}

\begin{figure}
    \centering
    \includegraphics[width=0.55\linewidth]{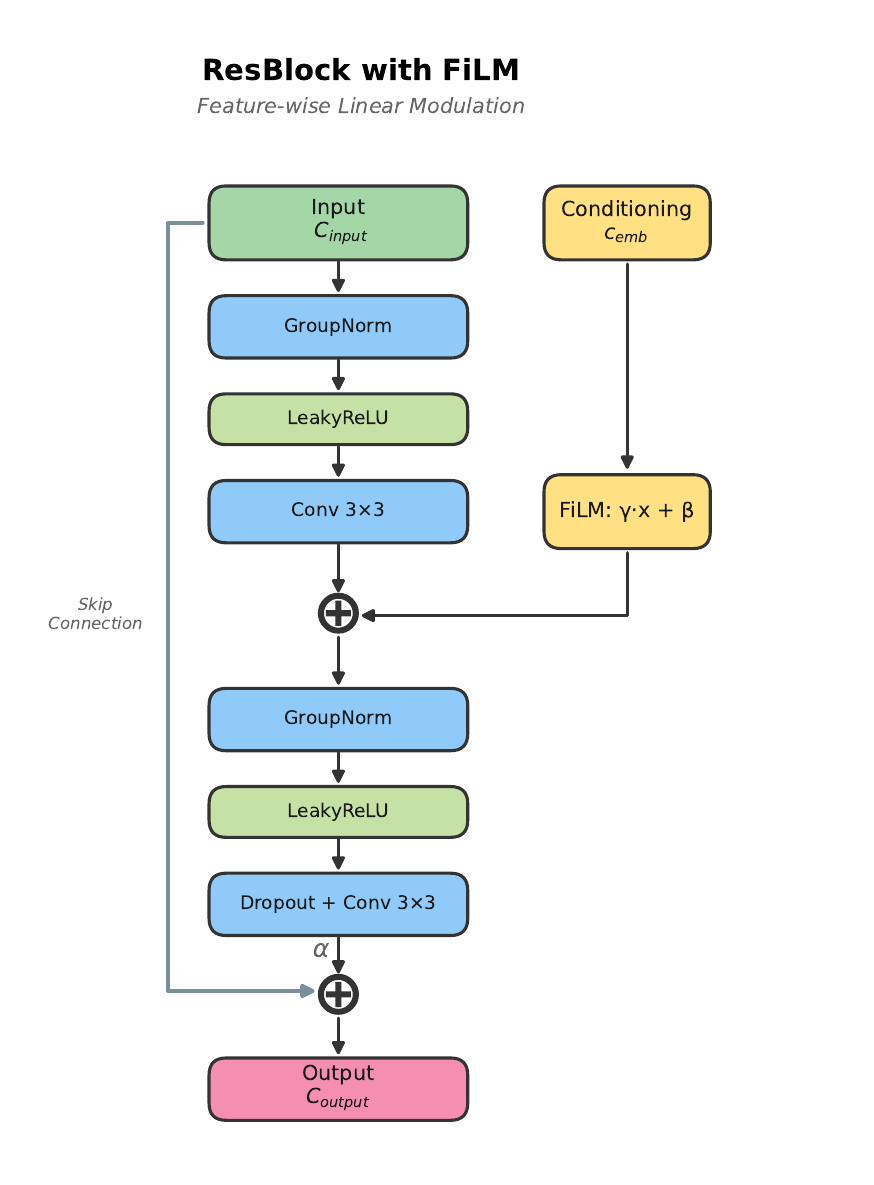}
    \caption{Schematic of the Residual Block with FiLM Conditioning. An additional convolution layer with kernel size $1$ is used for the skip connection if input channels $C_{input}$ differs from output channels $C_{output}$. The FiLM based conditioning outputs a $C_{output}$ vector as embedding, which is broadcast across the spatial dimensions. }
    \label{fig:resblock}
\end{figure}

The fundamental building block is a conditioned residual block. Let $h \in \R^{C_{\text{in}} \times H' \times W'}$ be the input feature map. The block applies two convolutional layers with normalization, activation, and FiLM conditioning
\begin{align*}
    h' &= \text{Conv}_{3\times3}^{(1)}(\sigma(\text{GroupNorm}_1(h))), \\
    h'' &= \text{FiLM}(h', \bm{e}), \\
    h''' &= \text{Dropout}_p\left(\text{Conv}_{3\times3}^{(2)}(\sigma(\text{GroupNorm}_2(h'')))\right), \\
    \text{ResBlock}(h, \bm{e}) &= \alpha \cdot h''' + \mathcal{R}(h),
\end{align*}
where $p$ is the dropout probability, $\alpha \in (0, 1]$ is a residual scaling factor that improves training stability for deep networks~\cite{szegedy2017inception}, and $\mathcal{R}$ is the residual projection
\begin{equation*}
    \mathcal{R}(h) = \begin{cases}
        h & \text{if } C_{\text{in}} = C_{\text{out}}, \\
        \text{Conv}_{1\times1}(h) & \text{otherwise}.
    \end{cases}
\end{equation*}
A schematic of the proposed residual block is shown in Figure \ref{fig:resblock}.
The scaling factor $\alpha < 1$ scales the residual block activations, which has been shown to facilitate optimization in very deep residual networks by preventing gradient explosion~\cite{szegedy2017inception}.

\subsubsection{U-Net Architecture}

The complete U-Net architecture consists of an encoder path $\text{Enc}_\theta$, a bottleneck $\text{Mid}_\theta$, and a decoder path $\text{Dec}_\theta$ with skip connections. The network is defined as
\begin{equation*}
    F_\theta(X_0, \bm{c}) = \text{Dec}_\theta \left( \text{Mid}_\theta \left( \text{Enc}_\theta(\tilde{X}_0, \bm{e}) \right), \bm{e} \right),
\end{equation*}
where $\bm{e} = \text{MLP}(\bm{c})$ is the conditioning embedding and $\tilde{X}_0$ is the coordinate-augmented input.

\paragraph{Encoder:} Starting from $h_0 = \text{Conv}_{\text{init}}(\tilde{X}_0) \in \R^{(C+2) \times H \times W}$, the encoder progressively reduces spatial resolution while increasing channel depth. For $\ell = 1, \ldots, L$,
\begin{equation*}
    h_\ell = \text{Down}_\ell\left(\text{ResBlock}_\ell^{\text{Enc}}(h_{\ell-1}, \bm{e})\right),
\end{equation*}
where $\text{Down}_\ell$ performs $2\times$ spatial downsampling via strided convolution, and $h_\ell \in \R^{C_0 m_\ell \times H/2^\ell \times W/2^\ell}$ with channel multiplier $m_\ell$. Skip features $\{h_0, h_1, \ldots, h_{L-1}\}$ are stored for the decoder.

\paragraph{Bottleneck:} At the coarsest resolution, two residual blocks process the encoded representation and the Attention mechanism as proposed by Vaswani et al.\cite{vaswani2017attention} is applied
\begin{equation*}
    z = \text{Attention}_{2}(\text{ResBlock}_2^{\text{Mid}}(\text{Attention}_{1}(\text{ResBlock}_1^{\text{Mid}}(h_L, \bm{e})), \bm{e})).
\end{equation*}

\paragraph{Decoder:} The decoder mirrors the encoder, progressively upsampling while fusing skip connections. For $\ell = L-1, \ldots, 0$,
\begin{equation*}
    \tilde{h}_\ell = \text{ResBlock}_\ell^{\text{Dec}}\left(\text{Up}_\ell(\tilde{h}_{\ell+1}) \oplus h_\ell, \bm{e}\right),
\end{equation*}
where $\text{Up}_\ell$ performs $2\times$ spatial nearest neighbor upsampling, and $\oplus$ denotes channel-wise concatenation.

\paragraph{Output:} The final prediction is obtained via
\begin{equation*}
    \hat{X}_M = \text{Conv}_{\text{final}}(\text{GroupNorm}(\tilde{h}_0)) \in \R^{C \times H \times W}.
\end{equation*}

\begin{algorithm}[H]
\caption{Conditioned U-Net Forward Pass}\label{alg:unet}
\begin{algorithmic}[1]
    \Require Initial field $X_0 \in \R^{C \times H \times W}$, parameter vector $\bm{c} \in \R^d$
    \Ensure Predicted solution field $\hat{X}_M$
    \State $\bm{e} \gets \text{MLP}(\bm{c})$ \Comment{Compute conditioning embedding}
    \State $\tilde{X}_0 \gets [X_0; \bm{\xi}; \bm{\eta}]$ \Comment{Augment with coordinate channels}
    \State $h_0 \gets \text{Conv}_{\text{init}}(\tilde{X}_0)$ \Comment{Initial convolution}
    \For{$\ell = 1$ to $L$} \Comment{\textbf{Encoder}}
        \State $h_\ell \gets \text{Down}_\ell(\text{ResBlock}_\ell^{\text{Enc}}(h_{\ell-1}, \bm{e}))$
    \EndFor
    \State $z \gets \text{Attention}_{2}(\text{ResBlock}_2^{\text{Mid}}(\text{Attention}_{1}(\text{ResBlock}_1^{\text{Mid}}(h_L, \bm{e})), \bm{e}))$ \Comment{\textbf{Bottleneck}}
    \State $\tilde{h}_L \gets z$
    \For{$\ell = L-1$ to $0$} \Comment{\textbf{Decoder}}
        \State $\tilde{h}_\ell \gets \text{ResBlock}_\ell^{\text{Dec}}(\text{Up}_\ell(\tilde{h}_{\ell+1}) \oplus h_\ell, \bm{e})$
    \EndFor
    \State $\hat{X}_M \gets \text{Conv}_{\text{final}}(\text{GroupNorm}(\tilde{h}_0))$ \Comment{Output projection}
    \State \Return $\hat{X}_M$
\end{algorithmic}
\end{algorithm}

\subsection{Neural Network Training}\label{sec:training}

\subsubsection{Huber Loss}

We employ the Huber loss~\cite{huber1964robust}, also known as smooth $\ell_1$ loss, as the primary training objective
\begin{equation}
    d_{\text{Huber}}(r) = \begin{cases}
        \frac{1}{2} r^2 & \text{if } |r| \leq \delta, \\
        \delta \left( |r| - \frac{1}{2}\delta \right) & \text{otherwise},
    \end{cases}
\end{equation}
and
\begin{equation}\label{eq:huberloss}
    \mathcal{L}_{\text{Huber}}(R) = \frac{1}{N^2}\sum_{i,j}^N d_{\text{Huber}}(r_{i,j}) 
\end{equation}
where $R = F_{\theta}(X_{0}^{(k)}, \mathbf{c}^{(k)}) - X_M$ is the residual, $r_{i,j}$ are the pixel-wise components of $R$, and $\delta = 1$ is the threshold parameter. The total loss over the dataset is
\begin{equation*}
    \mathcal{L}_{\mathcal{D}}(\theta) = \frac{1}{\mathcal{N}} \sum_{k=1}^{\mathcal{N}}\mathcal{L}_{\text{Huber}}\left( F_{\theta}(X_{0}^{(k)}, \mathbf{c}^{(k)}) - X_{M}^{(k)} \right).
\end{equation*}

The Huber loss combines the advantages of both $\ell_1$ and $\ell_2$ losses: it behaves quadratically for small residuals, providing smooth gradients near the optimum, while transitioning to linear behavior for large residuals, offering robustness to outliers~\cite{huber1964robust}. This hybrid behavior is particularly beneficial for PDE surrogate modeling, where solution fields may contain regions of varying smoothness and occasional sharp gradients.

\subsubsection{AdamW Optimizer}\label{sec:adamw}

Parameters are optimized using AdamW~\cite{loshchilov2019adamw}, which decouples weight decay from the adaptive learning rate mechanism. The update rule is
\begin{align*}
    m_t &= \beta_1 m_{t-1} + (1 - \beta_1) g_t, \\
    v_t &= \beta_2 v_{t-1} + (1 - \beta_2) g_t^2, \\
    \hat{m}_t &= \frac{m_t}{1 - \beta_1^t}, \quad \hat{v}_t = \frac{v_t}{1 - \beta_2^t}, \\
    \theta_{t+1} &= \theta_t - \eta \left( \frac{\hat{m}_t}{\sqrt{\hat{v}_t} + \epsilon} + \lambda_w \theta_t \right),
\end{align*}
where $g_t = \nabla_\theta \mathcal{L}(\theta_t)$ is the gradient, $m_t$ and $v_t$ are the first and second moment estimates, $\hat{m}_t$ and $\hat{v}_t$ are bias-corrected moments, $\eta$ is the learning rate, $\beta_1, \beta_2$ are exponential decay rates, $\epsilon$ is a small constant for numerical stability, and $\lambda_w$ is the weight decay coefficient.

Unlike the original Adam optimizer~\cite{kingma2015adam}, AdamW applies weight decay directly to the parameters rather than incorporating it into the gradient, which has been shown to improve generalization~\cite{loshchilov2019adamw}.

\subsubsection{Cosine Annealing Learning Rate Schedule}\label{sec:cosann}

The learning rate follows a cosine annealing schedule~\cite{loshchilov2017sgdr}
\begin{equation*}
    \eta_t = \eta_{\min} + \frac{1}{2}(\eta_0 - \eta_{\min})\left(1 + \cos\left(\frac{t \pi}{T_{\max}}\right)\right),
\end{equation*}
where $\eta_0$ is the initial learning rate, $\eta_{\min}$ is the minimum learning rate, $t$ is the current epoch, and $T_{\max}$ is the total number of epochs.

This schedule provides a smooth decay from $\eta_0$ to $\eta_{\min}$, with the rate of decrease being slower at the beginning and end of training, and faster in the middle. The cosine schedule has been shown to improve convergence compared to step-wise or linear decay schedules~\cite{loshchilov2017sgdr}.

\subsubsection{Gradient Clipping}\label{sec:gradclip}

To prevent gradient explosion and ensure stable training, we apply gradient clipping by norm~\cite{pascanu2013difficulty}
\begin{equation*}
    \tilde{g} = \begin{cases}
        \nabla_\theta \mathcal{L} & \text{if } \| \nabla_\theta \mathcal{L} \|_2 \leq \tau, \\
        \tau \cdot \frac{\nabla_\theta \mathcal{L}}{\|\nabla_\theta \mathcal{L}\|_2} & \text{otherwise},
    \end{cases}
\end{equation*}
where $\tau$ is the clipping threshold. This technique is particularly important for deep networks and when training on data with varying scales~\cite{pascanu2013difficulty}. 

\begin{table}[]
    \centering
    \begin{tabular}{ll}
        \toprule
        \textbf{Component} & \textbf{Specification} \\
        \midrule
        Input resolution & $H \times W = 64 \times 64$ \\
        Input/output channels & $C = 1$ \\
        Base channels & $C_0 = 32$ \\
        Channel multipliers & $(1, 2, 4, 8)$ \\
        Resolution levels & $L = 4$ \\
        Residual blocks per level & $N_r = 2$ \\
        Normalization & GroupNorm ($G = 8$)~\cite{wu2018groupnorm} \\
        Activation & LeakyReLU ($\alpha_{\text{neg}} = 0.01$)~\cite{maas2013leakyrelu} \\
        Conditioning dimension & $d = 4$, $d_e = 16$ \\
        Conditioning MLP multiplier & $\kappa = 2$ \\
        Residual scale & $\alpha = 0.2$ \\
        Dropout & $p = 0.05$ \\
        \midrule
        \textbf{Training} & \\
        \midrule
        Loss function & Huber ($\delta = 1$)~\cite{huber1964robust} \\
        Optimizer & AdamW~\cite{loshchilov2019adamw} \\
        Learning rate & $\eta_0 = 5 \times 10^{-4}$ \\
        Weight decay & $\lambda_w = 5 \times 10^{-3}$ \\
        Betas & $(\beta_1, \beta_2) = (0.9, 0.999)$ \\
        LR schedule & CosineAnnealing~\cite{loshchilov2017sgdr} \\
        Minimum LR & $\eta_{\min} = 10^{-6}$ \\
        Epochs & $T_{\max} = 100$ \\
        Gradient clip norm & $\tau = 1.0$ \\
        Batch size & $16$ \\
        \bottomrule
    \end{tabular}
    \caption{Architecture and training hyperparameters for the transport-diffusion surrogate model.}
    \label{tab:architecture}
\end{table}
\begin{figure}[!ht]
    \centering
    \includegraphics[width=\linewidth]{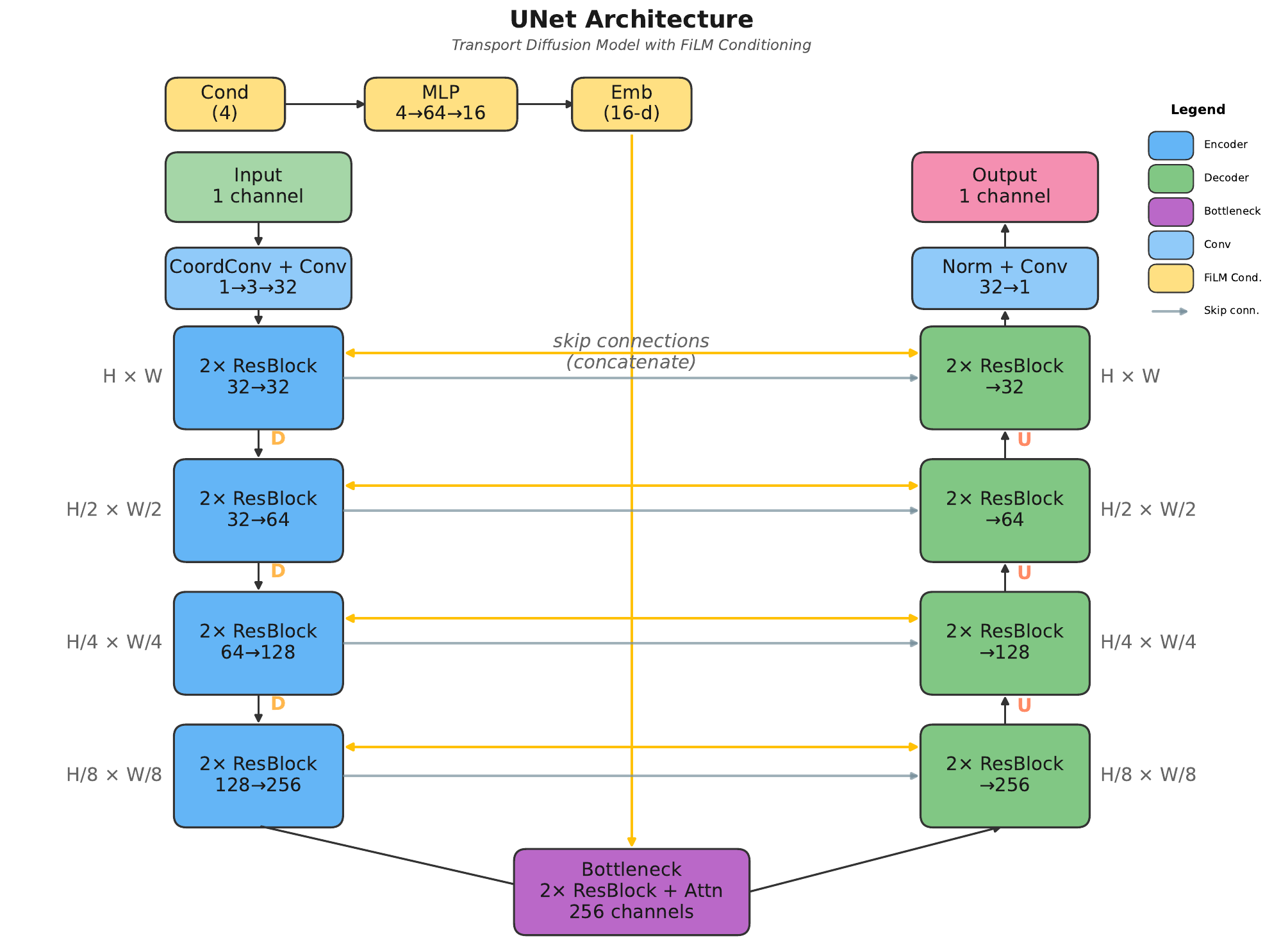}
    \caption{Schematic of the U-Net models architecture. The numbers in the second row of each block denotes the number of channels, while the light grey notations of $H \times W$ denote the changing spatial dimensions. $\mathbf{D}$ denotes a downsampling operation with factor $2$ and $\mathbf{U}$ denotes upsampling with a factor $2$ using the simple nearest neighbor infill.}
    \label{fig:unet-architecture}
\end{figure}
\subsection{Training Results}\label{sec:unetresults}
\begin{figure}[!ht]
    \centering
    \includegraphics[width=0.7\linewidth]{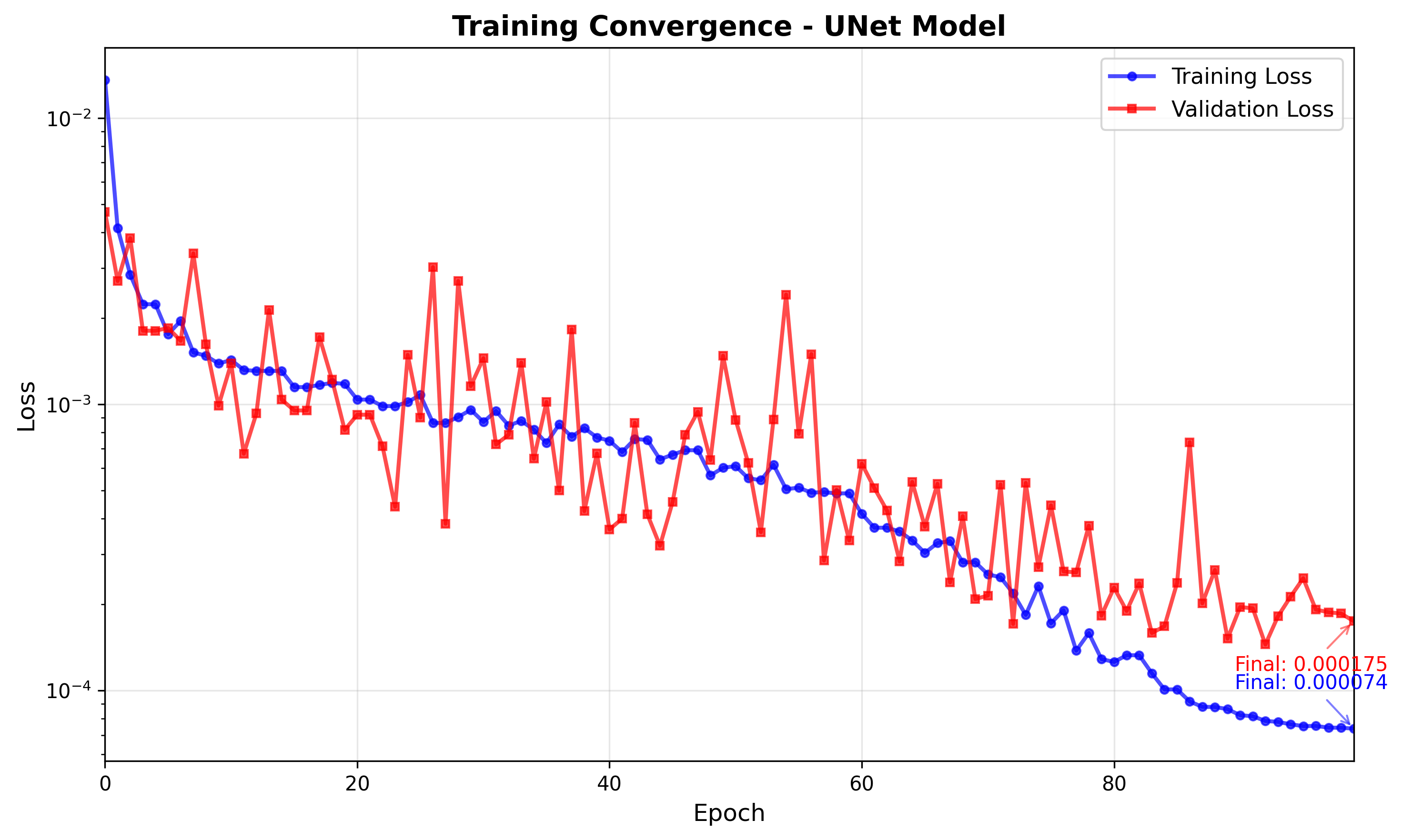}
    \caption{Convergence of the loss function over $100$ training epochs.}
    \label{fig:unetconvergence}
\end{figure}

\begin{figure}[!ht]
    \centering
    \includegraphics[width=0.95\linewidth]{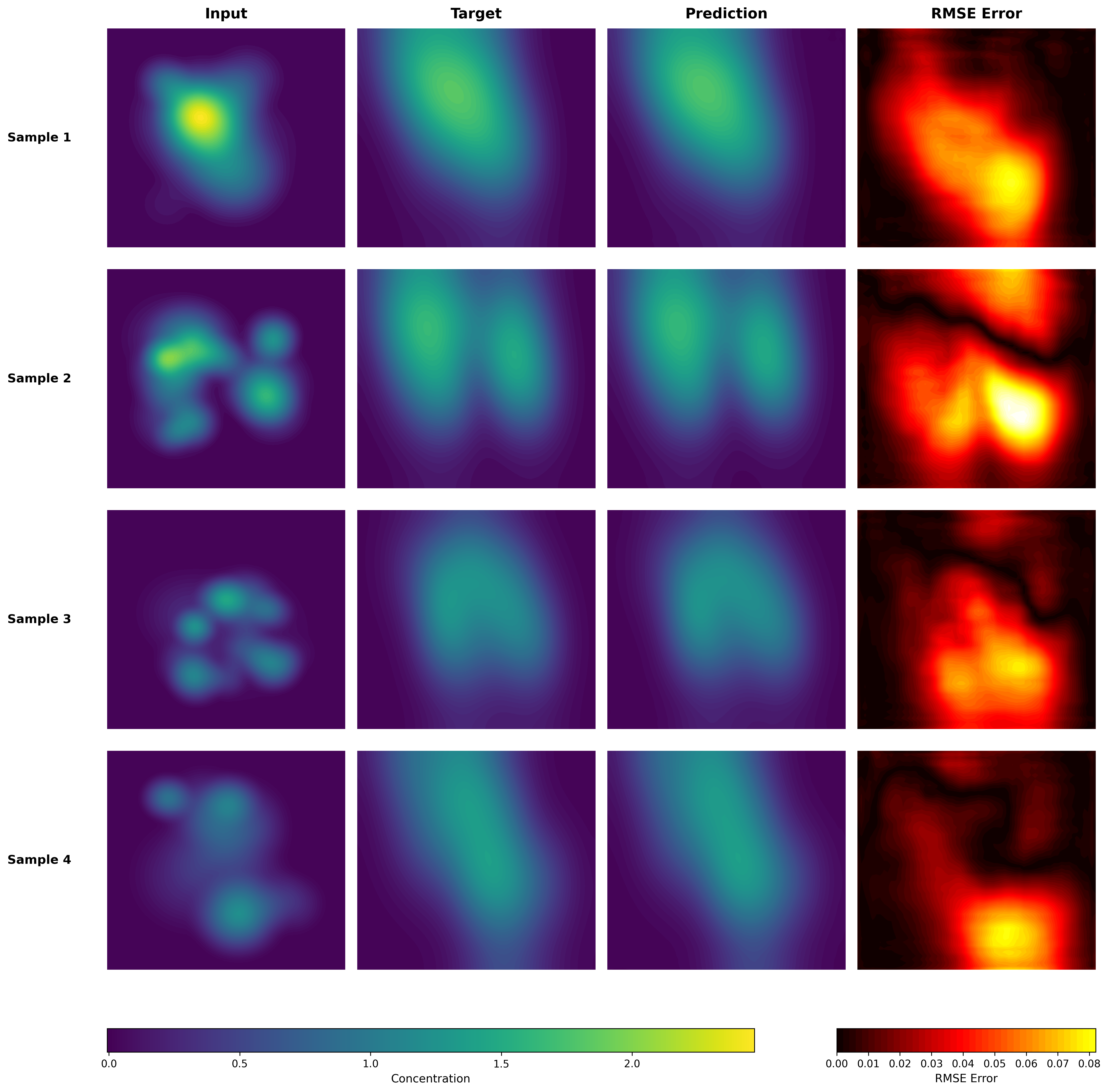}
    \caption{Image grid of test batch with model prediction and RMSE error visualization after training.}
    \label{fig:unetbatch}
\end{figure}

A conceptual visualization of the computational workflow of the model is shown in Figure \ref{fig:unet-architecture}.
The training convergence of the U-Net model under the conditions described in Table \ref{tab:architecture} is displayed in Figure \ref{fig:unetconvergence}.  The training loss function indicates convergence by flattening out after around $90$ epochs. The final training loss was $7.4 \times 10^{-5}$, while the validation loss is slightly above that with $17.5 \times 10^{-5}$. Around $70$ epochs, the training loss significantly falls below the validation loss. Both of these indicate the convergence of the training process. We show an exemplary batch prediction on test data, not seen by the model during the training process, in Figure \ref{fig:unetbatch}. The model predictions are visually indistinguishable, indicated by the maximum error of $0.08$ in row $2$ of the error plots. We further measured resource usage during the forward pass of the DNN model on the Elwetritsch cluster at RPTU. The results are shown in Table \ref{tab:unetruntime}. The runtime of the DNN prediction is effectively steady at $~11ms$ independent of the batch size. Therefore, the sample efficiency scales nicely with increasing batch size during inference. 

\begin{table}[H]
	\centering
	\begin{tabular}{c||c|c|c}
		batch size & total runtime [ms] & samples / s & peak memory usage [MB]\\[10pt]
		\hline \hline
		1 & 10.89 & 91.8 & 78.4  \\[10pt]
		2 & 10.79 & 185.4 & 81.3  \\[10pt]
		4 & 10.79 & 370.9 & 87.0  \\[10pt]
		8 & 10.83 & 738.4 & 107.6  \\[10pt]
        16 & 12.36 & 1294.7 & 154.0 \\[10pt]
	\end{tabular}
	\caption{Runtime and resource usage for batch prediction on Tesla V100-SXM2-32GB gpu.}
    \label{tab:unetruntime}
\end{table} 

\subsection{Model Generalization}\label{sec:modelgen}
The specific construction of the test dataset, as describe in Section \ref{subsec:Trainingdata}, allows a more thorough investigation into the models input mechanisms, initial condition $U_0$ and conditioning $\mathbf{c}$. 
We define the error matrix $E \in \mathbb{R}^{50 \times 50}$, with $i$ being the category of initial condition and $j$ being the category of the conditioning vector. 
Further, we define a scale-normalized Huber loss $\mathcal{L}_{\text{nHuber}}$ by normalizing residuals with the RMS magnitude $s(X_M)$. We replace $R$ in Equation \ref{eq:huberloss} with $R_n(X_{M}, X_{0}, \mathbf{c}) = \frac{F_{\theta}(X_{0}, \mathbf{c}) - X_M}{s(X_M) + \varepsilon}$, where $s(\cdot) =\vert \vert \cdot\vert\vert_{2} / \sqrt{CHW}$ and $\varepsilon=10^{-8}$ for numerical stability.
This yields a dimensionless error measure that allows comparisons across samples with varying overall density magnitude. Also, $\mathcal{L}_{\text{nHuber}}$ down weights large-amplitude fields less aggressively than per-pixel normalizations, which would focus on near-zero pixels predominately.

\begin{figure}[!ht] 
    \centering
    \includegraphics[width=0.7\linewidth]{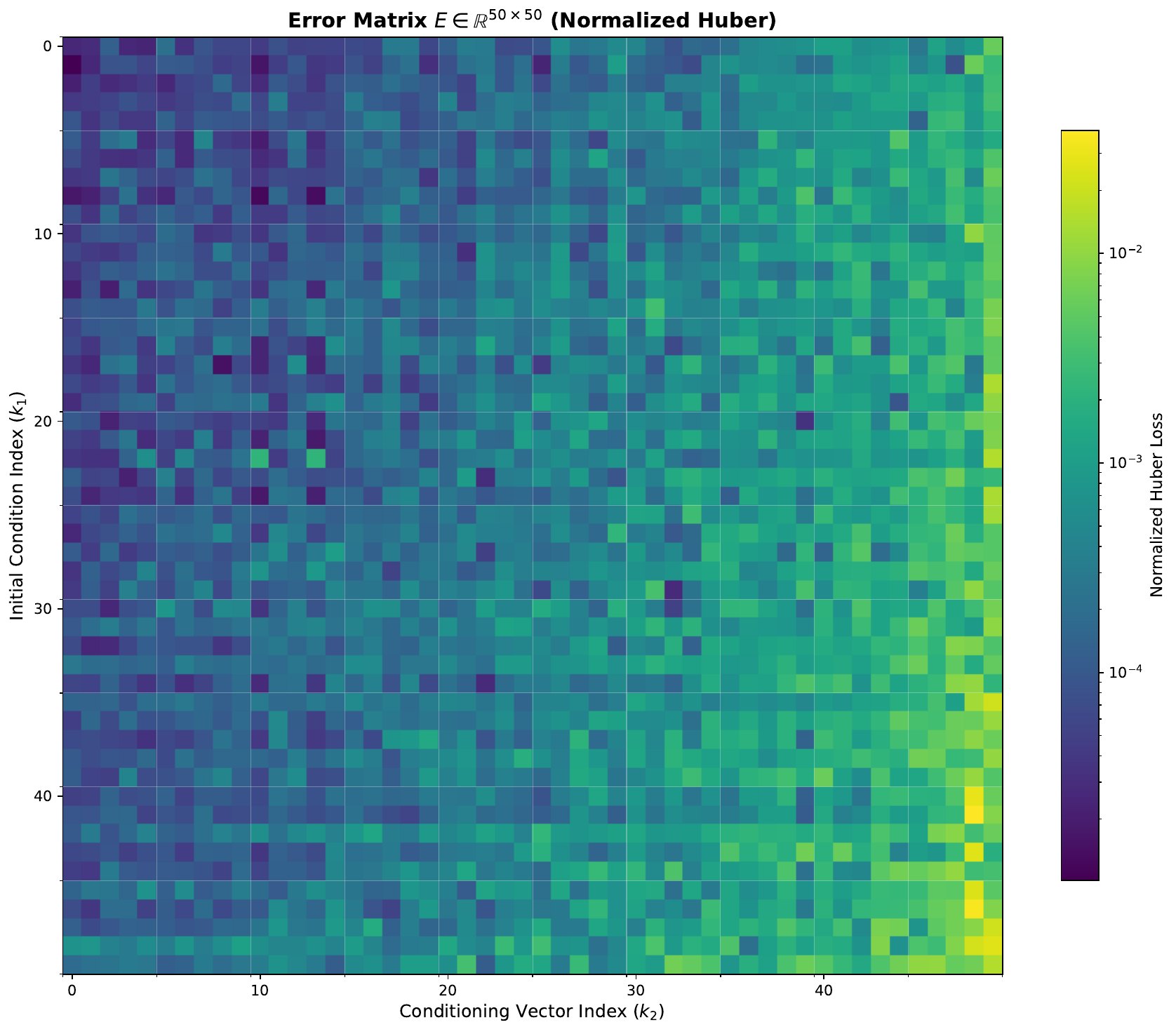}
    \caption{Error matrix $E \in \mathbb{R}^{50 \times 50}$ for the normalized Huber loss as heatmap with logarithmic scale and ordered by mean error along $k_1$ and $k_2$ axis.}
    \label{fig:error_matrix_heatmap}
\end{figure}

\begin{figure}[!ht]
    \centering
    \begin{subfigure}{0.49\textwidth}\label{fig:ic_grouping_huber}
        \includegraphics[width=\linewidth, trim={0 0 0 1.5cm},clip]{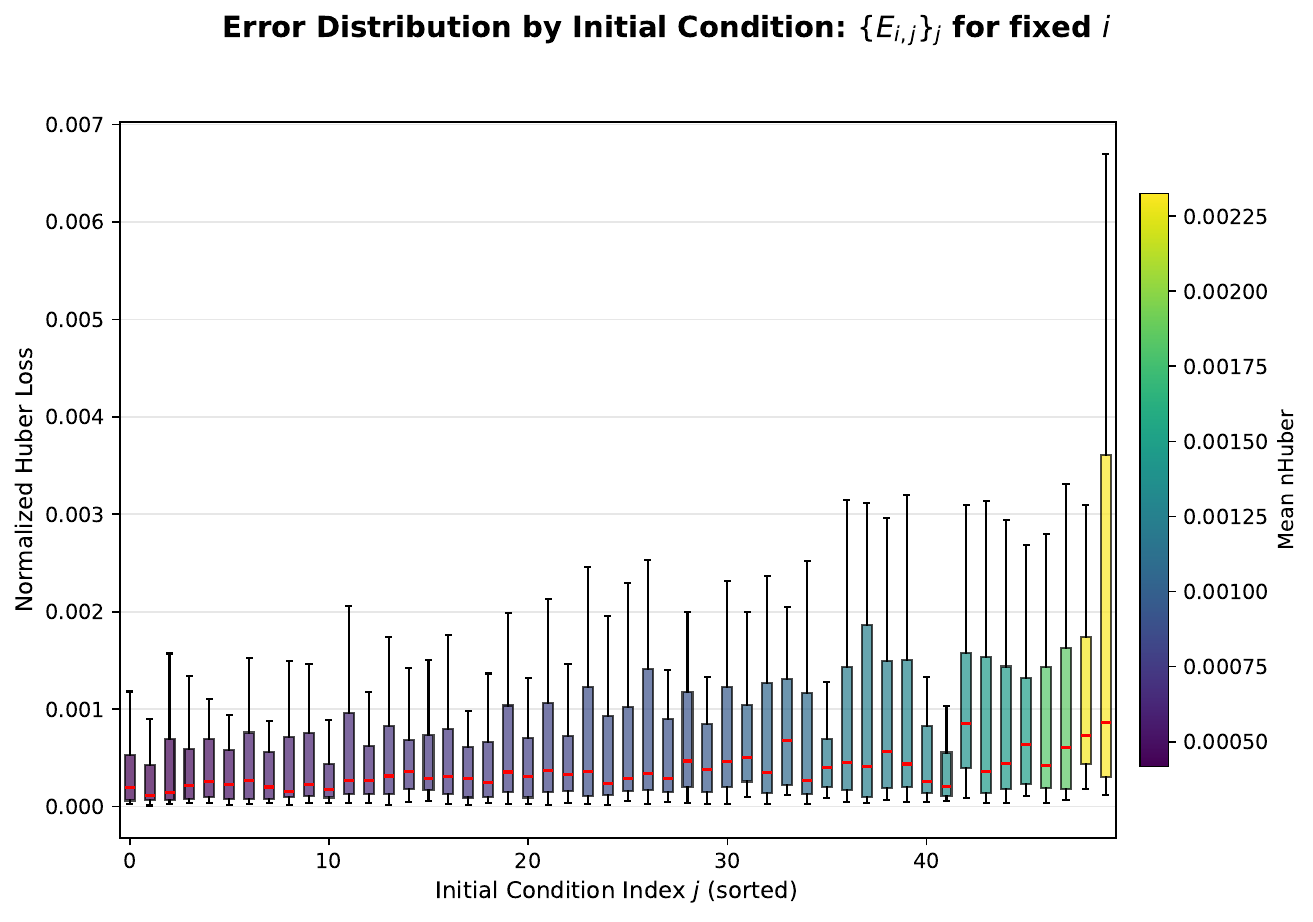}
        \caption{Error distribution by initial condition $\{ E_{i,j}\}_j$ for fixed $i$.}
         \label{fig:errors_grouped_a}
    \end{subfigure}
    \hfill
    \begin{subfigure}{0.49\textwidth}\label{fig:cond_grouping_huber}
        \includegraphics[width=\linewidth, trim={0 0 0 1.5cm}, clip]{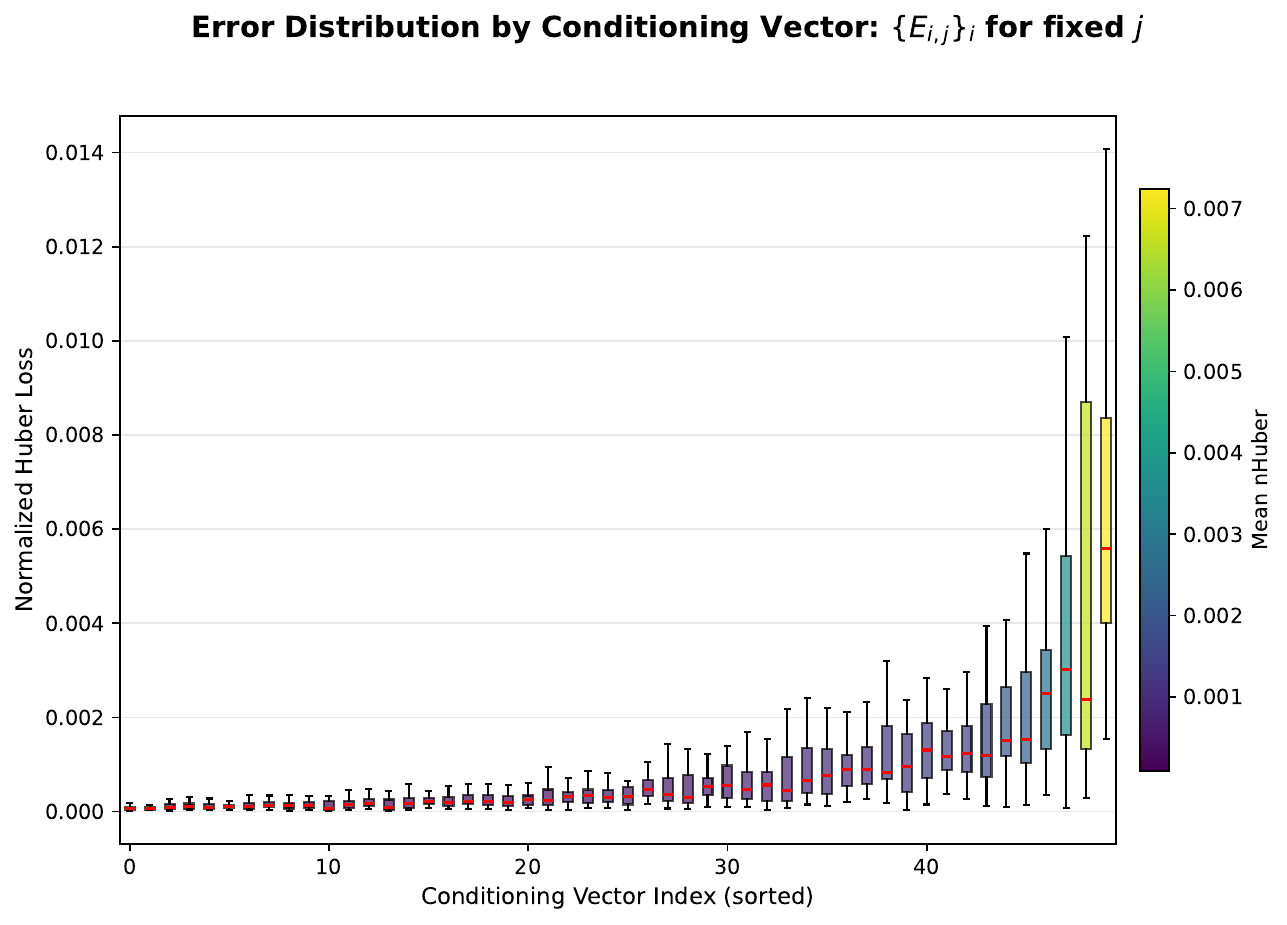}
        \caption{Error distribution by conditioning vector $\{ E_{i,j}\}_i$ for fixed $j$.}
         \label{fig:errors_grouped_b}
    \end{subfigure}
    \caption{Error distributions grouped by initial conditions and conditioning vector with removed outliers following Tukey's rule and sorted in ascending order by mean error.}
    \label{fig:errors_grouped}
\end{figure}

We visualize the error matrix as heatmap in Figure \ref{fig:error_matrix_heatmap} and order each 'category' by the mean across rows and columns respectively. We also show the grouped error distributions in Figure \ref{fig:errors_grouped}. 
Figure \ref{fig:errors_grouped} and \ref{fig:error_matrix_heatmap} indicate that variability is much larger across conditioning vector index $k_2$ than across initial condition index index $k_1$. The variance in $k_1$ is reasonably neglectable. However, we notice some trend considering the conditioning vector index $k_2$ and the $\mathcal{L}_{\text{nHuber}}$. In Figure \ref{fig:errors_grouped_b} we show the significant differences between the different conditioning vector groupings varying by more than a magnitude. In Table \ref{tab:error_rankings} we showcase the worst and best conditioning vectors of the test data ordered by the mean $\mathcal{L}_{\text{nHuber}}$ in each group.  
\begin{table}[!ht]
    \centering
    \begin{tabular}{c|c|c|c|c|c}
        rank & $c_1$ & $c_2$ & $c_3$ & $c_4$ & $\frac{1}{50}\mathcal{L}_{\text{nHuber}}$ $\downarrow$\\
        \hline
        50 & 0.083 & 2.018 & 4.501 & 1.174 & \color{red}{0.007241}\\
        49 & 3.926 & 1.740 & 7.724 & 2.007 & \color{red}{0.006630}\\
        48 & 1.004 & 1.072 & 3.537 & 1.945 & \color{red}{0.003655}\\
        47 & 5.021 & 2.754 & 4.342 & 1.863 & \color{red}{0.002780}\\
        46 & 3.097 & 0.402 & 1.067 & 1.002 & \color{red}{0.002244}\\
        \vdots & \vdots & \vdots & \vdots & \vdots & \vdots\\
        5 & 0.621 & -0.198 & 7.741 & 1.908 & \color{green}{0.000138}\\
        4 & 4.595 & 1.388 & 6.564 & 1.274 & \color{green}{0.000134}\\
        3 & 4.535 & 1.662 & 8.478 & 2.275 & \color{green}{0.000116} \\
        2 & 0.081 & 2.611 & 6.532 & 2.308 & \color{green}{0.000093} \\
        1 & 5.769 & -0.087 & 7.311 & 2.217 & \color{green}{0.000085} \\
    \end{tabular}
    \caption{Performance rankings for condition vectors (groupings) ordered by mean $\mathcal{L}_{\text{nHuber}}$ (descending) and condition vector values.}
    \label{tab:error_rankings}
\end{table}
An initial correlation based analysis with Pearson and Spearman correlation coefficients for correlations between conditioning dimension and the mean $\mathcal{L}_{\text{nHuber}}$ error is shown in Table \ref{tab:correlation_analysis}. 
\begin{table}[!ht]
    \centering
    \begin{tabular}{c|c|c}
        dim($\mathbf{c}$) & Pearson $r$ & Spearman $\rho$\\\hline
        1 & -0.0719 & 0.0206 \\
        2 & 0.2893 & 0.3097 \\
        3 & -0.1701 & -0.3657 \\
        4 &  -0.1185 & -0.1143 \\
    \end{tabular}
    \caption{Pearson and Spearman correlation coefficient for conditioning dimensions.}
    \label{tab:correlation_analysis}
\end{table}
Due to the nature of machine learning algorithms and models we argue there is a weak-to-moderate monotone correlation in dimension $2$ and a weak negative correlation in dimension $3$. This difficulty is likely governed by coefficient interactions and regime changes regarding the PDE characteristics. Further investigation is required to confirm the true reason. The findings only confirm that no single coefficient explains the difficulty. 

\section{Discussion  \& conclusions} \label{sec:conclusion}
 
In this study, we develop a robust and stable C++  numerical solver for a nonlinear, instationary convection-diffusion-reaction partial differential equation for generation of simulation-based data sets for the training and conditioning of the presented U-Net surrogate model. 
The implemented physical model can be generalized for the macroscopic modeling of cancer cell phenotypic dynamics. Our discretization scheme is based on the finite-difference method of second order as well as the third-order Runge–Kutta-Fehlenberg scheme as time discretization method.
Our numerical error study confirms that the numerical solution converges at a second-order rate to a highly resolved reference solution as the mesh size is refined, see Section \ref{subsec:convstudy}. 

Furthermore, time-step  control is implemented with an error tolerance of $10^{-6}$, which is lower than the minimum spatial discretization error achieved
(compare Fig. \ref{fig:L2erros} and Table \ref{tab:L2erros}). This guarantees the temporal discretization error to be negligible in comparison to the spatial error, as well as ensures the stability with respect to the time discretization step sizes. The latter can be indirectly observed in Table \ref{tab:rel_errors}, where the average time step gets approximately four times smaller when the mesh size $h$ is halved (with the exception of low-resolution meshes), indicating the quadratic dependence of the time step on the spatial step size, a condition that can be derived from a stability analysis, see e.g., \cite{hindmarsh}.

The mesh convergence study also provides useful information on the expected size of the discretization error and thus enables us to choose an appropriate mesh resolution when generating training and test data. As already mentioned in Section \ref{subsec:Trainingdata}, for the used mesh of $256\times 256$ grid points the expected relative $L^\infty$ errors imply that the maximal absolute error in each grid point is less than $0.1\%$ of the maximum of the reference solution. 

Four different coefficients are selected to parameterize our considered family of coefficient functions as indicated in Table \ref{tab:parameterfunctions}. In fact, for each of the different model mechanisms describing diffusion, convection and nonlinear population dynamics effects, there is at least one of the parameters $c_1,c_2,c_3,c_4$ randomized in order to develop a DNN model that can satisfactorily predict the influence of the variation of each model mechanism.
Moreover, the range of the allowed randomized coefficient values is chosen to balance the two  competing aspects: the numerical stability of the C++-solver and the impact of parameter variability on the solution predictions: If the parameter range is chosen too small, the influence of changes in this parameter on the solution will be negligible; if the parameter range is chosen too large, the parameters $c_1,c_2,c_3$ will cause stricter restrictions on the discretization step sizes, increasing the computational cost. In the worst case, if $c_3$ had been chosen outside of our chosen interval, the diffusion matrix $D$ may not be positive definit anymore.

\medskip

The conditioned U-Net surrogate introduced in Section \ref{sec:DNN} provides an accurate and computationally efficient approximation of the fixed-horizon solution map $(U_0, \mathbf{c})\mapsto U_M$ for the considered convection-diffusion-reaction dynamics. Under the training configuration summarized in Table \ref{tab:architecture}, optimization of the Huber objective (Equation \ref{eq:huberloss}) converges stably, with the learning curves leveling out after approximately $90$ epochs (compare Figure \ref{fig:unetconvergence}). Qualitative evaluation on held-out test data confirms that the surrogate captures the dominant spatial structures of the evolved density field. The predicted solutions in Figure \ref{fig:unetbatch} reproduce the spatial localization and transport-induced deformation patterns observed in the numerical reference data. Additionally, we showcased in Table \ref{tab:unetruntime} that the forward-pass runtime is essentially constant in batch inference, implying that large-scale parameter sweeps and repeated evaluations can be performed efficiently once the surrogate is trained. From a methodological perspective, the combination of Feature-wise Linear Modulation and coordinate encoding is essential. FiLM implements sample- and channel-wise affine modulation driven by the coefficient vector $\mathbf{c}$, enabling parameter sharing across regimes and the learning of a solution operator family, while coordinate channels mitigate the translation-invariance bias of standard convolutions and improve the representation of spatially inhomogeneous dynamics. 
Beyond the average predictive fidelity, the factorial construction of the test data set from Section \ref{subsec:Trainingdata} enables a structured assessment of generalization across initial conditions and parameter regimes. Through a scale-normalized formulation of the Huber metric introduced in Section \ref{sec:modelgen}, which allows a more scale independent analysis, we construct the error matrix $E \in \mathbb{R}^{50 \times 50}$, visualized in Figure \ref{fig:error_matrix_heatmap}. Combined with the grouped error distributions in Figure \ref{fig:errors_grouped}, the results show that error variability is larger across conditioning vectors than across initial conditions, indicating that approximation difficulty is governed primarily by the induced PDE regime rather than by specific initial density realizations within the sampled family. The rankings of conditioning vectors in Table \ref{tab:error_rankings} further emphasizes the vast difference in reconstruction quality by conditioning vector. The statistical analysis using Pearson/Spearman correlation coefficient in Table \ref{tab:correlation_analysis} reveals only weak-to-moderate monotone associations with individual coefficient dimensions. This suggests that surrogate error is not controlled by a single coefficient alone but emerges from interactions between transport, diffusion, and reaction effects and requires further analysis. Collectively, these findings support the central methodological claim that we can utilize the FiLM conditioned U-Net surrogate with coordinate encoding for a robust one-network-for-many-regimes surrogate that generalizes well across diverse initial fields and a prescribed coefficient range.

\medskip

\subsection{Outlook}\label{sec:outlook}

\medskip


The present study establishes a validated numerical reference solver and a parameter-conditioned U-Net surrogate for rapid prediction of convection–diffusion-reaction dynamics.
For a description of more complex macroscopic phenotypic plasticity mechanisms the considered PDE model could be generalized  to a PIDE model by including an additional integral term to model non-local changes of cell phenotypes as applied in \cite{chrisholm}. 
For similar reasons the boundary conditions can be adapted to  (mass conserving)  no-flux-condition or general Robin type boundary condition.

While the surrogate is trained for a fixed terminal time $T$ and a low-dimensional coefficient vector $\mathbf{c}$, the conditioning mechanism provides a natural pathway to inverse problems. Since $F_{\theta}(U_0, \mathbf{c})$ is differentiable with respect to both inputs, gradients $\nabla_\mathbf{c} F_{\theta}(U_0, \mathbf{c})$ are available
via automatic differentiation. This enable gradient-based parameter inference from observations $(U_{0}^{obs}, U_{M}^{obs})$ by solving optimization problems of the form
\begin{equation*}
    \min_{\mathbf{c} \in \mathcal{C}} \mathcal{J}(\mathbf{c}) = \mathcal{D}(F_{\theta}(U_0^{obs}, \mathbf{c}), U_{M}^{obs}) + \lambda \mathcal{R(\mathbf{c})},
\end{equation*}
where $\mathcal{D}$ is a discrepancy functional on the fields (e.g., relative $L^2$ or Huber) and $\mathcal{R}$ encodes prior knowledge or regularization, such as bounds consistent with the sampling ranges in Section $\ref{sec:initialcondtion}$. In these problem settings, the surrogate can serve as a fast inner-loop forward model, enabling rapid calibration, sensitivity analysis, and uncertainty quantification workflows that would be impractical with repeated numerical solves. The regime dependent generalization analysis in Section \ref{sec:modelgen} motivates several concrete extensions. First, the existence of systematically harder to learn conditioning regimes needs further investigation. \\

Second, the current conditioning could be generalized to using grid-wise parameter fields to improve the
generalizability and robustness of our physical P(I)DE model.
This would enable the use of model parameter functions $D,v,f$
defined by their discretized grid representations rather than a reduced parameterization, leading to a wider class of by the U-Net depictable parameter functions and extending FiLM-style modulation to higher-dimensional embeddings. \\

Finally, complementary to the surrogate-based gradients, algorithmic differentiation of the C++ solver (e.g. via CoDiPack \cite{sagebaum}) remains a principled option when exact discrete sensitivities are required. This approach has the advantage of circumventing the difficulties that arise in the classical adjoint methodology for sensitivity computation for complex, nonlinear time dependent problems. A pragmatic workflow would constitute the use of the surrogate for rapid screening and initialization, followed by refinement using solver-based sensitivities on a reduced set of candidates.\\

\section{Acknowledgment}
The support of this research by the Carl Zeiss Foundation (CZS) in the framework of the ‘CZS Breakthroughs’ programme under the joint project \href{https://www.carl-zeiss-stiftung.de/en/project-overview/detail/artificial-intelligence-for-treating-cancer-therapy-resistance-ai-care}{\sl AI Care}, and by  the research initiative of the federal state of Rhineland-Palatinate under the potential area \href{https://rptu.de/en/projects/mso}{\sl MSO} is gratefully acknowledged. Also, the authors gratefully acknowledge the funding of the German National High Performance Computing (NHR) association for the Center NHR South-West. 

\bibliographystyle{unsrt} 
\bibliography{references}

\end{document}